\documentclass[letterpaper]{article} 
\usepackage{aaai25}  
\usepackage{times}  
\usepackage{helvet}  
\usepackage{courier}  
\usepackage[hyphens]{url}  
\usepackage{graphicx} 
\usepackage{natbib}  
\usepackage{caption} 
\usepackage{algorithm}
\usepackage{algorithmic}

\usepackage{dsfont} 

\usepackage{algorithm}
\usepackage{algorithmic}
\usepackage{graphicx}
\usepackage{amsmath}
\usepackage{amssymb}
\usepackage[capitalize]{cleveref}
\usepackage{booktabs}
\usepackage{xcolor,colortbl}
\usepackage{color, colortbl}
\definecolor{tabhighlight}{HTML}{e5e5e5}
\definecolor{citecolor}{HTML}{0071bc}

\usepackage{subcaption}
\usepackage{sidecap}
\usepackage{multirow}
\usepackage{bm}

\usepackage{mathtools}
\usepackage{amsthm}
\theoremstyle{plain}
\newtheorem{theorem}{Theorem}[section]
\newtheorem{proposition}[theorem]{Proposition}

\theoremstyle{definition}

\theoremstyle{remark}

\usepackage{newfloat}
\usepackage{listings}
\DeclareCaptionStyle{ruled}{labelfont=normalfont,labelsep=colon,strut=off} 
\floatstyle{ruled}
\newfloat{listing}{tb}{lst}{}
\floatname{listing}{Listing}
\usepackage{balance}

\title{Learning Strategy Representation for Imitation Learning in Multi-Agent Games}
\author{
 Shiqi Lei\textsuperscript{\rm 1}\thanks{Equal contribution}, Kanghoon Lee\textsuperscript{\rm 2}\footnotemark[1], Linjing Li\textsuperscript{\rm 1,3}, Jinkyoo Park\textsuperscript{\rm 2}\thanks{Corresponding author}
}
\affiliations{
    \textsuperscript{\rm 1}Institute of Automation, Chinese Academy of Sciences (CASIA)\\
    \textsuperscript{\rm 2}Korea Advanced Institute of Science and Technology (KAIST)\\
    \textsuperscript{\rm 3}Beijing Wenge Technology Co., Ltd.\\
 leishiqi2022@ia.ac.cn, leehoon@kaist.ac.kr, linjing.li@ia.ac.cn, jinkyoo.park@kaist.ac.kr\\

}

\usepackage{bibentry}

\begin{document}

\maketitle

\begin{abstract}
The offline datasets for imitation learning (IL) in multi-agent games typically contain player trajectories exhibiting diverse strategies, which necessitate measures to prevent learning algorithms from acquiring undesirable behaviors.
Learning representations for these trajectories is an effective approach to depicting the strategies employed by each demonstrator. 
However, existing learning strategies often require player identification or rely on strong assumptions, which are not appropriate for multi-agent games.
Therefore, in this paper, we introduce the Strategy Representation for Imitation Learning (STRIL) framework, which (1) effectively learns strategy representations in multi-agent games, (2) estimates proposed indicators based on these representations, and (3) filters out sub-optimal data using the indicators.
STRIL is a plug-in method that can be integrated into existing IL algorithms.
We demonstrate the effectiveness of STRIL across competitive multi-agent scenarios, including Two-player Pong, Limit Texas Hold'em, and Connect Four.
Our approach successfully acquires strategy representations and indicators, thereby identifying dominant trajectories and significantly enhancing existing IL performance across these environments.
\end{abstract}

\section{Introduction}\label{sec:intro}

Although reinforcement learning has become a powerful technique for sequential decision-making in various domains such as robotic manipulation \cite{andrychowicz2020learning}, autonomous driving \cite{chen2019model}, and game playing \cite{vinyals2019grandmaster}, conventional reinforcement learning demands substantial online interactions with the environment, which can be costly and sample inefficient while potentially leading to safety risks \cite{berner2019dota, bojarski2016end}.
To address these issues, many methods have emerged to enable efficient learning using offline datasets generated by demonstrators. 
For example, imitation learning (IL) \cite{pomerleau1988alvinn} replicates actions from the offline dataset without reward information, while offline reinforcement learning \cite{fujimoto2019off, kumar2020conservative} is provided access to reward signals. 
Offline learning datasets are usually collected from multiple demonstrators to enlarge dataset scale and diversity \cite{sharma2018multiple,mandlekar2019scaling,mandlekar2021matters}, which leads to a dataset of behaviors with various characteristics. 
However, standard IL algorithms treat all data samples in the dataset as homogeneous, potentially learning undesired behaviors from sub-optimal trajectories. 

To address the above issue, the key insight in our proposed method is to assign each trajectory in the offline dataset with a unique learned attribute, i.e., strategy representation, so that we can further analyze each trajectory considering its specificity and filter out sub-optimal data.
With a precise depiction of each trajectory and their distribution on the representation space, we can judge the performance of each trajectory by only collecting a few (less than 5\%) data with trajectory rewards or even without any reward information.
In this work, we introduce Strategy Representation for Imitation Learning (STRIL), an efficient and interpretable approach designed to improve IL by filtering sub-optimal demonstrations from offline datasets. 

\begin{figure*}[t]
    \centering
    \includegraphics[width=0.80\textwidth]{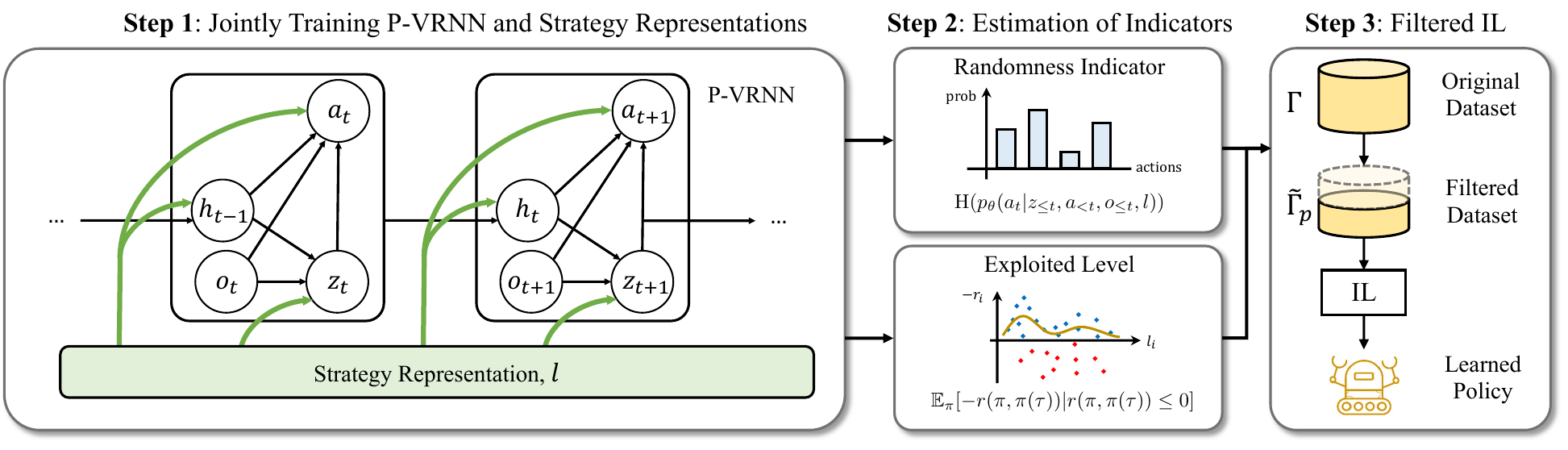}
    \vspace{-0.3cm}
    \caption{The overall diagram of Strategy Representation for Imitation Learning (STRIL).}
    \vspace{-0.3cm}
    \label{fig:overview}
\end{figure*}

\Cref{fig:overview} illustrates an overview of STRIL. Note that STRIL is a plug-in method compatible with existing IL algorithms. It consists of three components: strategy representation learning using a Partially-trainable-conditioned Variational Recurrent Neural Network (P-VRNN), indicator estimation, and filtered IL. The detailed steps and corresponding contributions are outlined as follows:

\begin{itemize}
    \item We propose an unsupervised framework with P-VRNN to efficiently extract strategy representations from multi-agent game trajectories. Strategy representation for each trajectory is customized as a network condition.
    \item We define the Randomness Indicator (RI) and Exploited Level (EL), which utilize strategy representation to effectively evaluate offline trajectories in a zero-sum game. EL can be precisely estimated even with limited reward data, while RI requires no reward data.
    \item We enhance existing IL methods by filtering out sub-optimal trajectories using the RI and EL indicators, ensuring that IL is trained only on the dominant trajectory.
    \item We demonstrate that STRIL can provide effective strategy learning without player identification and significantly enhance the performance of various IL algorithms in competitive zero-sum games, including Two-player Pong, Limit Texas Hold’em, and Connect Four.
\end{itemize}

\section{Related Works}
\textbf{Imitation Learning.}
In the conventional IL settings, the expert trajectories only have information of state-action pairs without reward information \cite{pomerleau1988alvinn, ross2011reduction, ho2016generative, ding2019goal,garg2021iq}, and it is assumed that the demonstrations are homogeneous oracle. 
However, realistic crowd-sourced datasets are usually multi-modal and include sub-optimal demonstrations. Some IL methods are proposed for multi-modal offline datasets, such as \cite{hausman2017multi} and \cite{fei2020triple}. As for sub-optimal data, there are plenty of approaches to alleviate the negative influence \cite{brown2020better, chen2021learning, zhang2021confidence, kim2022demodice, xu2022discriminator}, but all these methods require environment dynamics, the rankings over the demonstrations, or the identification of demonstrators. In contrast, our method does not require such information. 
\citet{sasaki2020behavioral} enhances behavior cloning (BC) with noisy demonstrations, but their method does not deal with general sub-optimal trajectories. TRAIL \cite{yang2021trail} achieves sample-efficient IL via a learned latent action space and a factored transition model. We would like to additionally mention the work by \citet{franzmeyer2024select}, which adopts a similar framework of filtering the offline dataset and uses an IL algorithm. Nevertheless, their method assumes a cooperative setting and requires reward information. 

\textbf{IL with Representation Learning.} 
The work by \citet{beliaev2022imitation} closely aligns with our research, sharing the primary goal of extracting expertise levels of trajectories.
They assume that the demonstrator has a vector indicating the expertise of latent skills, with each skill requiring a different level at a specific state. These elements jointly derive the expertise level.
The method also considers the policy worse when it is closer to uniformly random distribution.
However, this assumption cannot be satisfied even in simple games like RPS, where a uniformly random strategy constitutes a Nash equilibrium.
Play-LMP \cite{lynch2020learning} leverages unsupervised representation learning in a latent plan space for improved task generalization. However, employing a variational auto-encoder (VAE) with the encoder outputting latent plans is unsuitable for multi-player games, potentially leaking opponent information from the observations and disrupting the evaluation of the demonstrator.
\citet{grover2018learning} also studies learning policy representations from offline trajectories. They use the information of agent identification during training, which enables them to add a loss to distinguish one agent from others. However, this information may not be provided in the offline datasets.

\section{Preliminaries}

\textbf{Markov Games.} A Markov game \cite{littman1994markov} is a partially observable Markov decision process \cite{kaelbling1998planning} (POMDP) adapted to a multi-agent setting, where each agent has its own reward function. 
In a Markov game, there is a state space $S$ and $n$ agents, with each agent $i$ having a corresponding action space $A_i$ and observation space $O_i$.
When an agent is not required to take action at a certain state, its action space contains only one action, referred to as a `null' action.
At each time step $t$: (1) each agent $i$ obtains an observation $o_t^{i}\in O_i$ and selects an action $a_t^{i}\in A_i$ based on the policy of agent $i$, $\pi_i:O_i\times A_i \to [0,1]$; (2) the agent receives a reward $\mathfrak{r}_t^{i}:S\times A_i \to \mathbb{R}$ based on the state and the action; (3) the state is changed according to the transition function $T:S\times A_1 \times ... \times A_n \to S$. For a complete trajectory $\tau=((o_0^i,a_0^i),...,(o_T^i,a_T^i))$ of agent $i$, there is a reward $\mathfrak{r}_t^i$ received at each time step $t$. We define the trajectory reward $\tau$ as $\hat{r}_i(\tau)=\sum_{t=0}^T \mathfrak{r}_t^i$. The expected trajectory reward of player $i$ with strategy $\pi_i$ and opponents with strategy $\pi_{-i}$ is defined as $r_i(\pi_{-i},\pi_i)= \mathbb{E}_{\tau\sim (\pi_{-i},\pi_i)}\left[\hat{r}_i(\tau)\right]$, in which $\tau\sim(\pi_{-i},\pi_i)$ denote that $\tau$ is generated with agents using strategy $(\pi_{-i},\pi_i)$.

\begin{figure*}[ht]
  \centering
  \includegraphics[width=0.8\textwidth]{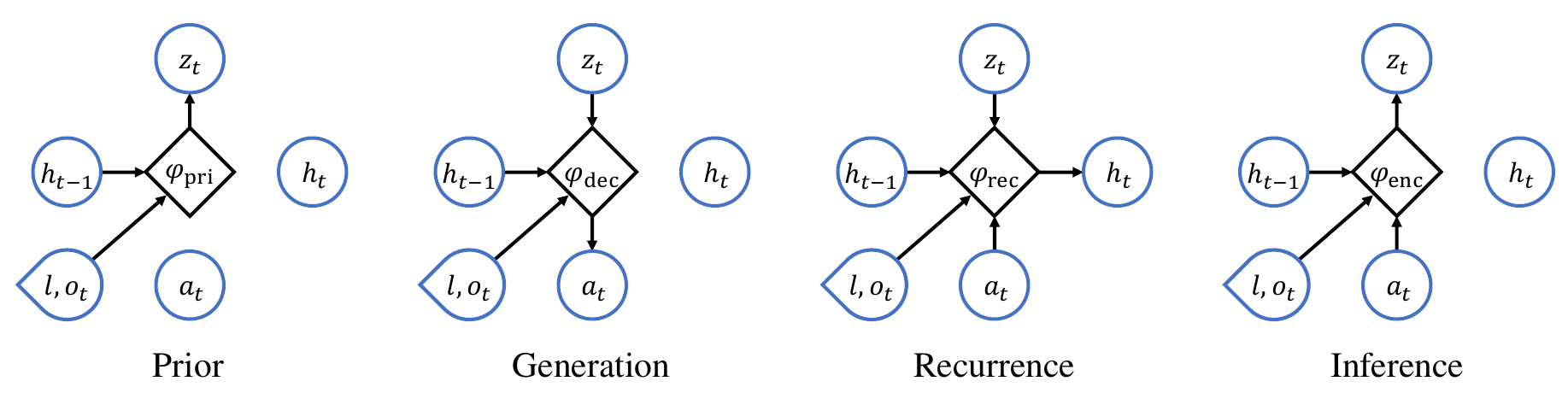}
  \caption{The decomposed network structure of the P-VRNN model. The variables are depicted as circles, learnable parameters as diamonds, and partially-trainable variables as a combination of both diamonds and circles.}
  \vspace{-0.2cm}
  \label{fig:network}
\end{figure*}

\textbf{Best Response, Exploitability, and Nash Equilibrium.} We use $r_i(\pi_{-i},\pi_i)$ to specify the reward of the player playing $\pi_i$ against $\pi_{-i}$. The best response of opponent strategy $\pi_{-i}$ is defined as $BR(\pi_{-i})=\text{argmax}_{\pi_i'}r_i(\pi_{-i}, \pi_i')$, which refers to the strategy of player $i$ that maximizes player $i$’s reward. 
We additionally define the best response of strategy $\pi_i$ as 
$BR(\pi_i)=\text{argmax}_{\pi_{-i}'}\sum_{j\in P,j\neq i}r_j(\pi_{-i}',\pi_i)$,
which equals to $\text{argmin}_{\pi_{-i}'}r_i(\pi_{-i}',\pi_i)$ in the zero-sum case. 
$BR(\pi_i)$ refers to the strategy of all the other players except player $i$ that maximizes their trajectory reward, which is equivalent to minimizing player $i$’s payoff in zero-sum games.
Let $P$ be the set of all the players in the game, and the strategy $\pi=(\pi_i)_{i\in P}$ be the strategy of all the players. 
We define the exploitability of strategy $\pi$ as 
$E(\pi)=\sum_{i\in P}\left(r_i(\pi_{-i},BR(\pi_{-i}))-r_i(\pi_{-i},\pi_i)\right)$, which reflects the extent to which the strategy can be exploited.
In zero-sum symmetric cases, we define the exploitability of a player strategy $\pi_i$ as 
$E(\pi_i)=-r_i(BR(\pi_{i}),\pi_i) =\sum_{j\in P,j\neq i}r_j(BR(\pi_{i}), \pi_i)$. A strategy $\pi_i$ is $\varepsilon$-Nash equilibrium if $E(\pi_i)\leq\varepsilon$.

\section{Problem Formulation}

Consider a multi-player competitive zero-sum game, and we have a dataset of game histories that include the trajectories of each player. The trajectories are generated by diverse players, ranging from high-level experts to amateurs. We aim to extract strategy representations from trajectories, distinguish the players with different levels, and learn an expert policy from the dataset via imitation learning. We assume that we do not have the identifications of the players.
In our problem, we collect a set $\Gamma$ of trajectories $\tau=((o_0,a_0),...,(o_T,a_T))$ from different games and demonstrators. The trajectory reward for a subset $\Gamma'\subset\Gamma$ is available for exploited level estimation. We assume that the strategy of a player is consistent within a single trajectory.

\section{Learning Strategy Representation}

Identifying the strategy of a player is essential to evaluating their skill level. However, this becomes challenging when player identification is unavailable in the dataset because the strategy of the player changes according to their opponent within each episode. 
Therefore, we propose a Partially-trainable-conditioned Variational Recurrent Neural Network (P-VRNN) featuring a strategy representation that is learnable and remains constant throughout the trajectory.
Strategy representation becomes the optimal representation for each trajectory by training it to minimize the P-VRNN loss.

The P-VRNN models the player's decision-making process and includes four major components similar to the original VRNN, as shown in \Cref{fig:network}. To disentangle the strategy of the opponent player from the strategy representation, we consider the observation as a conditional variable. 
We define $p$ as the generative model and $q$ as the inference model.

\subsection{Generation}

Based on the dependency of our P-VRNN model, We can model the decision-making process as follows:

\begin{equation}\label{eq:generation}
\begin{aligned}
    p(&a_{\leq T}, z_{\leq T}|o_{\leq T}, l) \\ &=\prod_{t=1}^{T} \underbrace{p(a_t|z_{\leq t}, a_{< t}, o_{\leq t}, l)}_{\text{generation}}\underbrace{p(z_t|a_{<t}, z_{<t}, o_{\leq t}, l)}_{\text{prior}},
\end{aligned}
\end{equation}

Without knowing the action $a_t$, the prior distribution of latent variable $z_t$ can be derived from the past actions $a_{<t}$, past latent variables $z_{<t}$, observations $o_{\leq t}$, and strategy representation $l$.
The computation graph of the P-VRNN shows that the recurrent variable $h_{t-1}$ integrates the past actions $a_{<t}$, latent variables $z_{<t}$, and observations $o_{<t}$. 
Therefore, with the assumption of a Gaussian distribution for the prior, the sampling process of $z_t$ is influenced by $h_{t-1}$, the current observation $o_t$, and the strategy representation $l$ as follows:
\begin{equation}
\begin{aligned}
&z_t\mid h_{t-1}, o_t, l\sim \mathcal{N}(\mu_{\text{pri}, t}, \text{diag}(\sigma_{\text{pri}, t}^2)),\\
&[\mu_{\text{pri}, t}, \sigma_{\text{pri}, t}] = \varphi_{\text{pri}}(h_{t-1}, o_t, l),
\end{aligned}
\end{equation}
where $\varphi_{\text{pri}}$ is a prior network. We also follow the convention in VAE and assume that the latent variable has a diagonal covariance matrix.

The generation process is obtaining action $a_t$ from the latent variables $z_{\leq t}$, past actions $a_{<t}$, observations $o_{\leq t}$, and strategy representation $l$ just same as to the decision-making process of the player. By substituting the past information using recurrent variable $h_{t-1}$, action generation is defined as follows:
\begin{equation}
\begin{aligned}
&a_t\mid h_{t-1}, z_t, o_t, l \sim \text{Cat}(\mu_{\text{dec}, t}),\\
&\mu_{\text{dec}, t} = \varphi_{\text{dec}}(h_{t-1}, z_t, o_t, l),
\end{aligned}
\end{equation}
where $\text{Cat}$ stands for categorical distribution and $\varphi_{\text{dec}}$ is a decoder network.

The recurrent unit takes in all the variables of the current step and the recurrent variable of the previous step, which includes all the past information. At each time step, $h_t$ is updated as follows:
\begin{equation}
h_t=\varphi_{\text{rec}}(h_{t-1}, a_t, z_t, o_t, l),
\end{equation}
where $\varphi_{\text{rec}}$ is a recurrent network.

\subsection{Inference}
Approximate posterior inference is modeled as follows:
\begin{equation}\label{eq:inference}
\begin{aligned}
    q(z_{\leq T}|a_{\leq T}, o_{\leq T}, l) = \prod_{t=1}^{T}\underbrace{q(z_t|a_{\leq t},z_{<t}, o_{\leq t}, l)}_{\text{inference}}.
\end{aligned}
\end{equation}
The latent variable $z_t$ is obtained from the actions $a_{\leq t}$, past latent variables $z_{<t}$, observations $o_{\leq t}$, and strategy representation $l$. Like the prior and action generation, we replace the past information with a recurrent variable of the previous step, $h_{t-1}$. Therefore, the approximate
posterior distribution is defined as follows:
\begin{equation}
\begin{aligned}
&z_t\mid h_{t-1}, a_t, o_t, l \sim \mathcal{N}(\mu_{\text{enc}, t}, \text{diag}((\sigma_{\text{enc}, t})^2)),\\
&[\mu_{\text{enc}, t}, \sigma_{\text{enc}, t}]=\varphi_{\text{enc}}(h_{t-1}, a_t, o_t, l),
\end{aligned}
\end{equation}
where $\varphi_{\text{enc}}$ is a encoder network.

\subsection{Learning}

Similar to \cite{chung2015recurrent}, the loss function of P-VRNN is a negative of the variational lower bound, using \Cref{eq:generation,eq:inference}, as follows:
\begin{equation}\label{eq:loss}
\begin{aligned}
\mathcal{L}=\mathbb{E}_{q_\phi(z_{\leq T}|a_{\leq T},o_{\leq T}, l)}\left[\sum_{t=1}^{T}\left(\mathcal{L}_{\text{Recon},t}+\mathcal{L}_{\text{Reg},t}\right)\right].
\end{aligned}
\end{equation}
The reconstruction loss for each timestep, which evaluates how well the generated action aligns with the original action, is formulated as follows:
\begin{equation}
\mathcal{L}_{\text{Recon},t} = -\log p_{\theta}(a_t|z_{\leq t}, a_{< t}, o_{\leq t}, l).
\end{equation}
The regularization loss for each timestep, which measures the divergence between the posterior and prior distributions, is formulated as follows:
\begin{equation}
\begin{aligned}
&\mathcal{L}_{\text{Reg},t} \\&= \mathrm{KL}(q_\phi(z_t|a_{\leq t},z_{<t}, o_{\leq t}, l) \!\parallel\! p_\theta(z_t|a_{<t}, z_{<t}, o_{\leq t}, l))
\end{aligned}
\end{equation}

At the beginning of the training, the strategy representation $l$, which is a trainable variable, is randomly initialized for each trajectory $\tau$.
The condition part of P-VRNN consists of an observation $o_t$ that changes over time and the strategy representation $l$, which is consistent during the whole trajectory and trainable.
During the training, all the $l$ are optimized together with the parameters of 
$\varphi_{\text{pri}}$, $\varphi_{\text{enc}}$, $\varphi_{\text{dec}}$, and $\varphi_{\text{rec}}$ to minimize the loss function described in \Cref{eq:loss}.
While the networks are trained to minimize the loss across all trajectories, the strategy representations are individually optimized for each trajectory to provide customized guidance and insights. Consequently, the strategy representation $l$ should be adjusted to more effectively capture and express the unique strategies of each trajectory. It is important to note that the process of deriving $l$ is conducted unsupervised, without needing player identification, ensuring privacy and generalizability.

\section{Indicators for Imitation Learning}

Utilizing the learned P-VRNN and the strategy representation for each trajectory, we propose the RI and EL indicators.

\subsection{Randomness Indicator (RI)}

Given the well-trained P-VRNN and the strategy representation dataset, we can evaluate the reconstruction loss and regularization loss for each trajectory. The regularization loss shows the capability of the posterior to approximate the prior, which reflects the performance of extracting the information of the next action from the past information, observation, and strategy representation. 
In the process of P-VRNN training, the regularization loss of each trajectory is gradually optimized to a very small value close to $0$. However, the reconstruction loss typically cannot be so small since the action decoder gives a probability distribution over actions, and players usually do not act deterministically. 
For a well-trained P-VRNN, the predicted action distribution closely matches the true probability distribution of the corresponding strategy of the trajectory. So if there are $n$ possible actions $a_{t,1}, a_{t,2},..., a_{t,n_t}$ for $a_t$, we can approximately calculate the expectation of the one-step reconstruction loss as
\begin{equation}
\begin{aligned}
&\mathbb{E}_{p_{\theta}(a_t|z_{\leq t}, a_{< t}, o_{\leq t}, l)}\left[\mathcal{L}_{\text{Recon},t}\right] \\
&=\sum_{i=1}^{n_t}-p_{\theta}(a_{t,i}|z_{\leq t}, a_{< t}, o_{\leq t}, l)\log p_{\theta}(a_{t,i}|z_{\leq t}, a_{< t}, o_{\leq t}, l) \\
&=\mathrm{H}\left(p_{\theta}(a_t|z_{\leq t}, a_{< t}, o_{\leq t}, l)\right).
\end{aligned}
\end{equation}
It is the entropy of $p_{\theta}(a_t|z_{\leq t}, a_{< t}, o_{\leq t}, l)$, which reflects the randomness of the player with strategy representation $l$, given $z_{\leq t}, a_{< t}$, and $o_{\leq t}$. Following the hypothesis of \citet{beliaev2022imitation}, a strategy with more randomness is considered worse.
Since we have the whole trajectory with a unified strategy representation $l$, we can define the RI of a trajectory as its cumulative reconstruction loss:
\begin{equation}
RI(\tau)=\sum_{t=1}^T \mathrm{H}\left(p_{\theta}(a_t|z_{\leq t}, a_{< t}, o_{\leq t}, l)\right).
\end{equation}
We highlight that the RI does not require any reward information, and the whole procedure is fully unsupervised.

\subsection{Exploited Level (EL)}\label{subsec:EL}

If we can access the trajectory rewards of select trajectories, we can determine to what extent the strategy of each trajectory in the offline dataset is exploited by utilizing the geometric structure of the strategy representation space.
The key insight of this approach is that the trajectories with similar strategy representations tend to exhibit similar strategies.

We define measure $\text{d}\pi$ on the strategy space $\Pi$, and assume that the strategy of agents generating dataset $\Gamma$ and its subset $\Gamma'$ are both sampled according to $\text{d}\pi$. 
Denote a trajectory as $\tau$ and the representation function mapping trajectories to learned representations as $f(\tau)$.
We remark that a trajectory $\tau$ should be mapped to a probability distribution of strategies such that $\int_{\pi\in\Pi}\tau(\pi)\text{d}\pi=1$, where $\tau(\pi)$ is the probability of using strategy $\pi$ when having trajectory $\tau$, instead of a single strategy. But we can view the mixture of $\pi$ with probability $\tau(\pi)$ as a single mixed strategy $\int_{\pi\in\Pi}\pi\tau(\pi)\text{d}\pi$, so we can still use notation $\pi(\tau)$ to represent the strategy of $\tau$. We define the EL as follows:
\begin{align}
    EL(\tau) &= \mathbb{E}_{\pi}\left[-r(\pi,\pi(\tau))\mid r(\pi,\pi(\tau))\leq 0\right] \\
    &= \frac{\int_{\pi\in\Pi}(-r(\pi,\pi(\tau))^{+}\text{d}\pi}{\int_{\pi\in\Pi}\mathds{1}_{r(\pi,\pi(\tau))\leq 0}\text{d}\pi},
\end{align}
where $r(\pi,\pi(\tau))$ returns the expected trajectory reward of a player with strategy $\pi(\tau)$ by default, $(x)^+=\max{\{x,0\}}$, and $\mathds{1}_{c}=1$ if and only if condition $c$ is satisfied, otherwise $\mathds{1}_{c}=0$. 
$EL(\tau)$ is the negative of the expectation of the trajectory reward less than $0$ when played with the demonstrators who generate the offline dataset. This value can reflect the extent to which the demonstrators exploit the strategy of $\tau$. 
To estimate EL with latent representation space structure, we provide an alternative definition of $EL_\delta$: 
\begin{equation}
EL_\delta(\tau)=\frac{\sum_{d(f(\tau),f(\tau'))<\delta}(-r(\hat{\pi}, \pi(\tau')))^+}{\sum_{d(f(\tau),f(\tau'))<\delta}\mathds{1}_{r(\hat{\pi}, \pi(\tau'))\leq 0}},
\end{equation}
where $d$ is a metric over the strategy representation space. 
Due to the Lipschitz continuity of the P-VRNN with respect to the representation, the trajectories with similar strategy representations have similar strategies. Thus, to approximate the negative EL of $\tau$, we can calculate the mean of all the negative rewards of the trajectories with the strategy representations in the small neighborhood of $\tau$’s representation. 
It can be proved that $\lim_{\delta\to 0^{+}}EL_\delta(\tau)=EL(\tau)$. $EL_\delta$ has favorable properties, such as the low value near Nash equilibrium strategies. Given a trajectory $\tau$ and its corresponding distribution $\tau(\pi)$ over $\Pi$, $\pi(\tau)$ is $\epsilon_1$-Nash equilibrium, and we assume that any pure strategy can exploit another strategy by at most $M$. 
We also assume that similar representations induce similar strategies: if $d(f(\tau_1),f(\tau_2))<\delta$, then $\int_{\pi\in\Pi}|\tau_1(\pi)-\tau_2(\pi)|\text{d}\pi<\alpha\delta$, where $\alpha$ is a constant.
It can be proved that $EL_\delta(\tau)<\epsilon_1+\alpha\delta M$.

Since EL is the average of values satisfying conditions with distance constraints on the representation space, we can train an operator $L$ to estimate EL from representation. We have the representation $l$ and trajectory reward $\hat{r}$ for each trajectory $\tau$, and we intend to minimize $\sum_{\hat{r}^i\geq 0}||L(l^i)-\hat{r}^i||_2$, where $l^i$ and $\hat{r}^i$ are the representation and trajectory reward of the $i$-th trajectory $\tau^i$ in the dataset, so that the prediction from $L(l)$ becomes close to the mean of satisfying reward $\hat{r}\geq 0$ nearby. We use a two-layer MLP as $L$. 
After training $L$, we can directly obtain the EL of a single trajectory $\tau$ even without the reward information. 
By applying EL estimator $L$ on the representation $f(\tau)$ of trajectory $\tau$, we can get the desired result $L(f(\tau))$.

\section{Filtered Imitation Learning}

The last step of the STRIL is to filter the offline dataset with a chosen percentile $p$ of an indicator. The indicator $I$ can be any mapping from the trajectories to real numbers such as RI or EL. Specifically, for an indicator $I(\tau)$, the offline dataset $\Gamma$ is filtered into $\Tilde{\Gamma}_p=\{\tau\in\Gamma\mid I(\tau)<I_p\}$, where $I_p$ satisfies that $\mathbb{P}_{\tau}[I(\tau)<I_p]=p$. After filtering the dataset, the original IL algorithm is employed. For IL algorithms that directly define loss function over target function and trajectories, the new loss function can be explicitly written as 
\begin{equation}
\mathcal{L}_p(\pi)=\mathbb{E}_{\tau}\left[\mathds{1}_{I(\tau)< I_p}\cdot\mathcal{L^{\text{IL}}}(\pi, \tau)\right],
\end{equation}
where $\mathcal{L}^{\text{IL}}(\pi,\tau)$ is the loss function of the IL algorithm. As an example, $\mathcal{L}^{\text{IL}}(\pi,\tau)=\sum_{t=0}^{|\tau|}\log \pi(a_t\mid o_t)$ in vanilla BC algorithm. 
As the value of $p$ closer to $0$, more data is filtered out; conversely, setting $p$ to 1 filters none of the data, reverting STRIL to the original IL algorithm.

\section{Experiments}\label{sec:exp}
\subsection{Experiment Settings}\label{sec:exp_setting}
We validate our approach using two-player zero-sum games: Two-player Pong, Limit Texas Hold’em \cite{zha2020rlcard}, and Connect Four \cite{terry2021pettingzoo}. 

\begin{figure*}[t]
  \centering
  \subfloat[Player ID]
  {
    \label{fig:pong1}\includegraphics[width=0.22\textwidth]{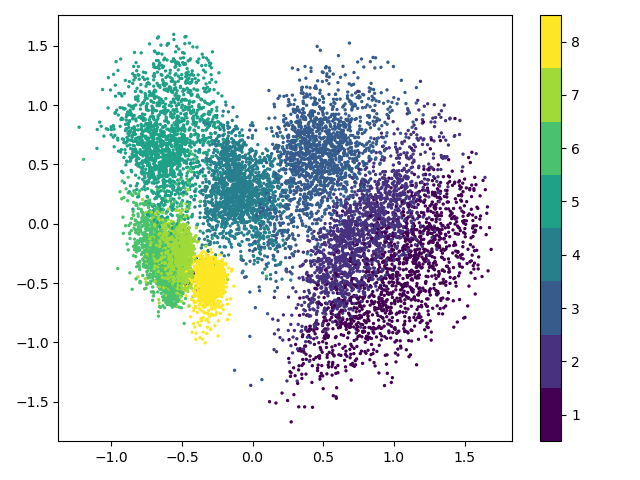}
  }
  \subfloat[Randomness indicator]
  {
    \label{fig:pong2}\includegraphics[width=0.22\textwidth]{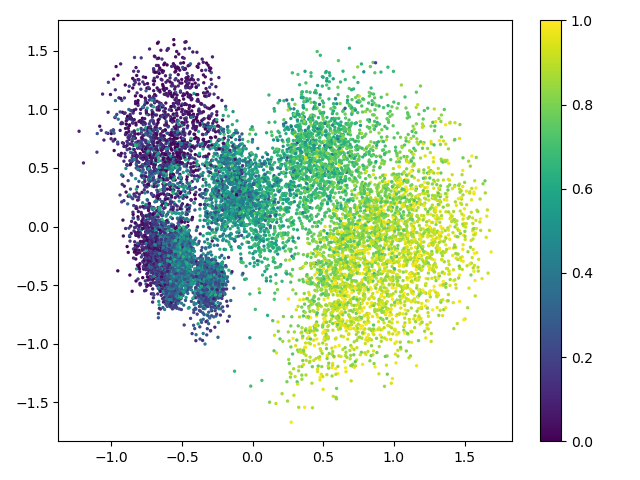}
  }
  \subfloat[Exploited level]
  {
    \label{fig:pong3}\includegraphics[width=0.22\textwidth]{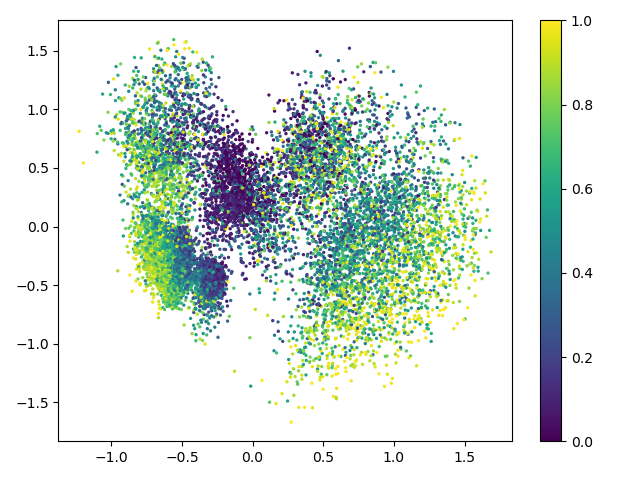}
  }
  \subfloat[Trajectory reward]
  {
    \label{fig:pong4}\includegraphics[width=0.22\textwidth]{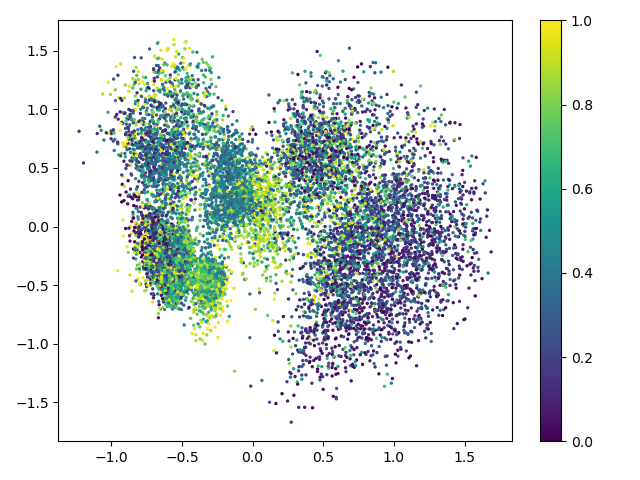}
  }
  \\
  
  \subfloat[Player ID]
  {
    \label{fig:card1}\includegraphics[width=0.22\textwidth]{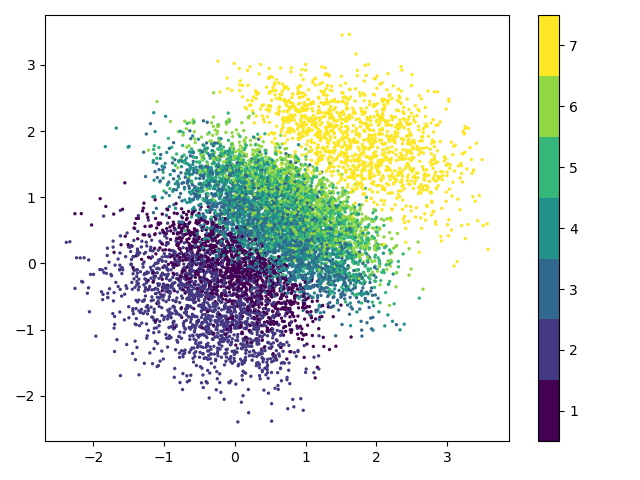}
  }
  \subfloat[Randomness indicator]
  {
    \label{fig:card2}\includegraphics[width=0.22\textwidth]{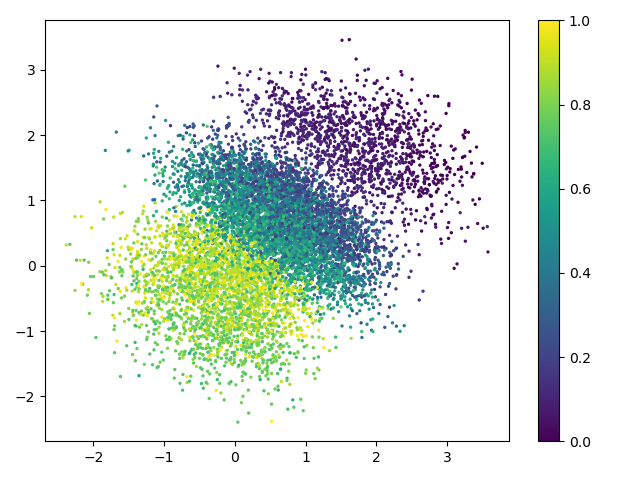}
  }
  \subfloat[Exploited level]
  {
    \label{fig:card3}\includegraphics[width=0.22\textwidth]{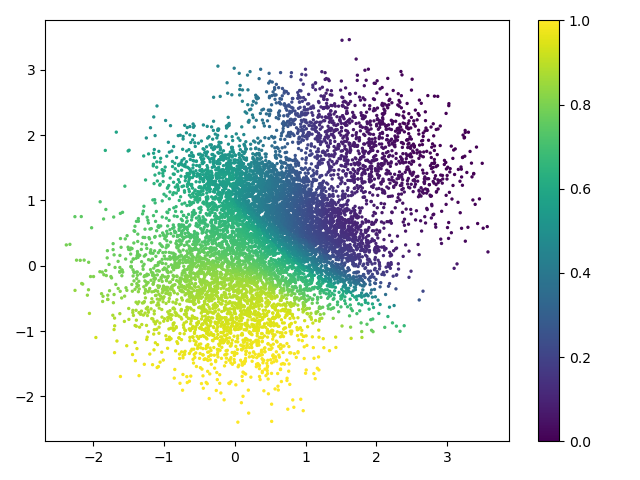}
  }
  \subfloat[Trajectory reward]
  {
    \label{fig:card4}\includegraphics[width=0.22\textwidth]{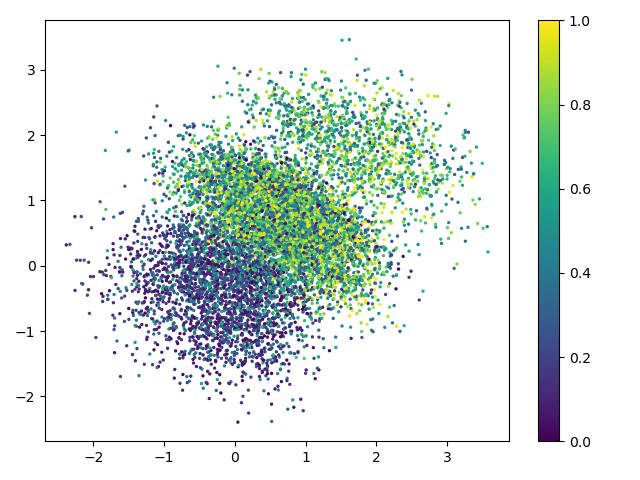}
  }\\
  
  \subfloat[Player ID]
  {
    \label{fig:conn1}\includegraphics[width=0.22\textwidth]{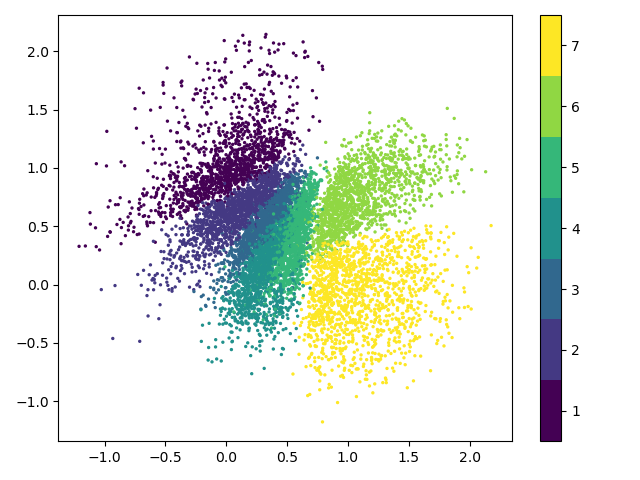}
  }
  \subfloat[Randomness Indicator]
  {
    \label{fig:conn2}\includegraphics[width=0.22\textwidth]{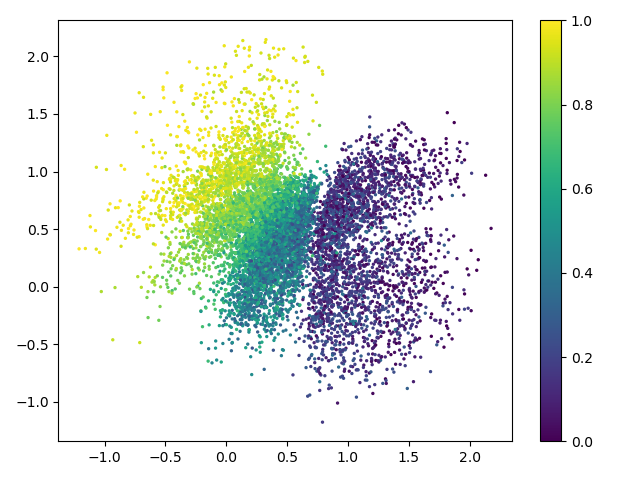}
  }
  \subfloat[Exploited Level]
  {
    \label{fig:conn3}\includegraphics[width=0.22\textwidth]{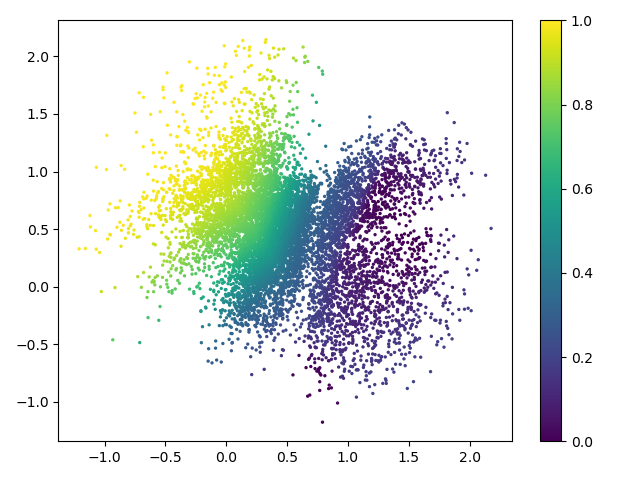}
  }
  \subfloat[Trajectory reward]
  {
    \label{fig:conn4}\includegraphics[width=0.22\textwidth]{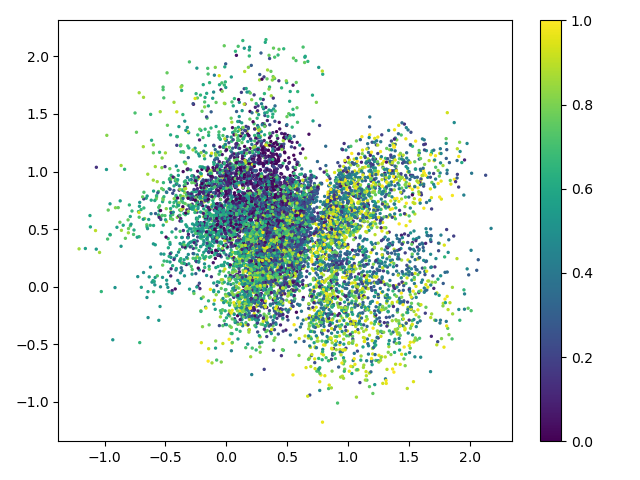}
  }\\
  
  \caption{The learned strategy representations with different labels on the Two-player Pong {(a-d)}, Limit Texas Hold’em {(e-h)}, and Connect Four {(i-l)} environments.}
  \vspace{-0.5cm}
  \label{fig:rps}
\end{figure*}

\textbf{Dataset generation.} We employ different methods to create training datasets with diverse demonstrators for the environments.
For Two-player Pong and Connect Four, we use self-play with opponent sampling \cite{bansal2018emergent} with the Proximal Policy Optimization (PPO) algorithm \cite{schulman2017proximal}.
For Limit Texas Hold'em, we use neural fictitious self-play \cite{heinrich2016deep} with Deep Q-network (DQN) algorithm \cite{mnih2013playing} to generate expert policies, given its complexity and the need to adapt to various opponents. 
Behavior models are then selected from multiple intermediate checkpoints to generate the offline data. We assume that only 5\% of the dataset is reward-labeled for EL estimation.

\textbf{Evaluation metrics.}
We evaluate our method across three environments to demonstrate the effectiveness of the learned strategy representation in STRIL using estimated indicators.
In a zero-sum game, evaluating policy performance involves ensuring the policy is not vulnerable to exploitation by a specific strategy.
In order to capture the worst-case scenario against opponent strategies in the dataset, we evaluate the performance of the imitative model, $\pi_i$, using the Worst Score ($\operatorname{WS}$) over the demonstrator set, $\mathcal{I}$:
\begin{equation}
\operatorname{WS}(\mathcal{I}, \pi_i)=\min_{j\in \mathcal{I}}r_i(\pi_j, \pi_i),
\end{equation}
where ${r}_i(\pi_j, \pi_i)$ represents the trajectory reward of $i$ against $j$.
For Two-player Pong and Connect Four, we calculate the reward using the formula $(N_{\text{win}} - N_{\text{lose}}) / N_{\text{game}}$, where $N_{\text{win}}$, $N_{\text{lose}}$, and $N_{\text{game}}$ represent the number of wins, losses, and total games, respectively.
For Limit Texas Hold’em, where a player can win by varying margins depending on the game, we determine the reward as the average difference between the total chips won and lost. We set $N_{\text{game}}$ to 2,000.

\subsection{Strategy Representation with Indicators}
In this subsection, we visualize the learned strategy representations using multiple labels. If the latent space exceeds two dimensions, it is initially reduced to two dimensions using PCA. These reduced representations are then color-coded based on different labels: player ID, RI, EL, and trajectory reward.
Note that player ID and trajectory reward serve as ground truth references while RI and EL are estimated. Instead of using the exact values, we color the percentiles of RI, EL, and trajectory reward.

\begin{table*}[h]
\centering
\begin{tabular}{c@{\hspace{12pt}}c@{\hspace{12pt}}c@{\hspace{12pt}}c@{\hspace{12pt}}c@{\hspace{12pt}}c}
\toprule
\multirow{2}{*}{\textbf{Game}} & \multirow{2}{*}{\textbf{Algorithm}} & \multicolumn{4}{c}{\textbf{Filtering Method}} \\
& & \textbf{Original} & \textbf{RI} & \textbf{EL} & \textbf{Best} \\
\midrule
\multirow{3}{*}{\textbf{Two-Player Pong}} 
& BC & $-0.832\pm0.011$ & $-0.613\pm0.052$ & \bm{$-0.343\pm0.036$} & \underline{$-0.044\pm0.033$} \\
& IQ-Learn & $-0.804\pm0.044$ & $-0.601\pm0.008$ & \bm{$-0.254\pm0.063$} & \underline{$-0.009\pm0.013$} \\
& ILEED & $-0.711\pm0.070$ & $-0.607\pm0.058$ & \bm{$-0.458\pm0.118$} & \underline{$-0.031\pm0.016$} \\
\midrule
\multirow{3}{*}{\textbf{Limit Texas Hold'em}} 
& BC & $-1.255\pm0.123$ & $0.532\pm0.052$ & \bm{$0.662\pm0.011$} & $0.464\pm0.103$ \\
& IQ-Learn & $-3.652\pm0.428$ & \bm{$0.667\pm0.097$} & $0.618\pm0.027$ & $0.640\pm0.061$ \\
& ILEED & $-0.411\pm0.150$ & \bm{$0.654\pm0.033$} & $0.494\pm0.065$ & $0.487\pm0.058$ \\
\midrule
\multirow{3}{*}{\textbf{Connect Four}} 
& BC & $-0.353\pm0.119$ & $0.255\pm0.080$ & \bm{$0.471\pm0.082$} & $0.407\pm0.053$ \\
& IQ-Learn & $-0.246\pm0.138$ & $0.117\pm0.131$ & \bm{$0.332\pm0.035$} & \underline{$0.393\pm0.047$} \\
& ILEED & $0.250\pm0.034$ & \bm{$0.267\pm0.081$} & $0.203\pm0.060$ & $0.005\pm0.219$ \\
\bottomrule
\end{tabular}
\vspace{-0.1cm}
\caption{$\operatorname{WS}$ of each IL algorithm over the demonstrator set, $\mathcal{I}$. Each algorithm was trained with distinct datasets filtered by various methods: (1) \textbf{Original}: utilizing the full dataset for training; (2) \textbf{RI}: filtering the dataset using the RI indicator; (3) \textbf{EL}: filtering the dataset using the EL indicator; and (4) \textbf{Best}: employing only the data generated by the dominant demonstrator, which serves as an oracle method. 
Bold highlights the best performance among Original, RI, and EL, while underline shows if the 'Best' method achieves the highest performance overall.
Higher is better.}
\vspace{-0.2cm}
\label{tab:exploitability}\end{table*}

\begin{figure*}[t]
  \centering
  \includegraphics[width=0.9\textwidth]{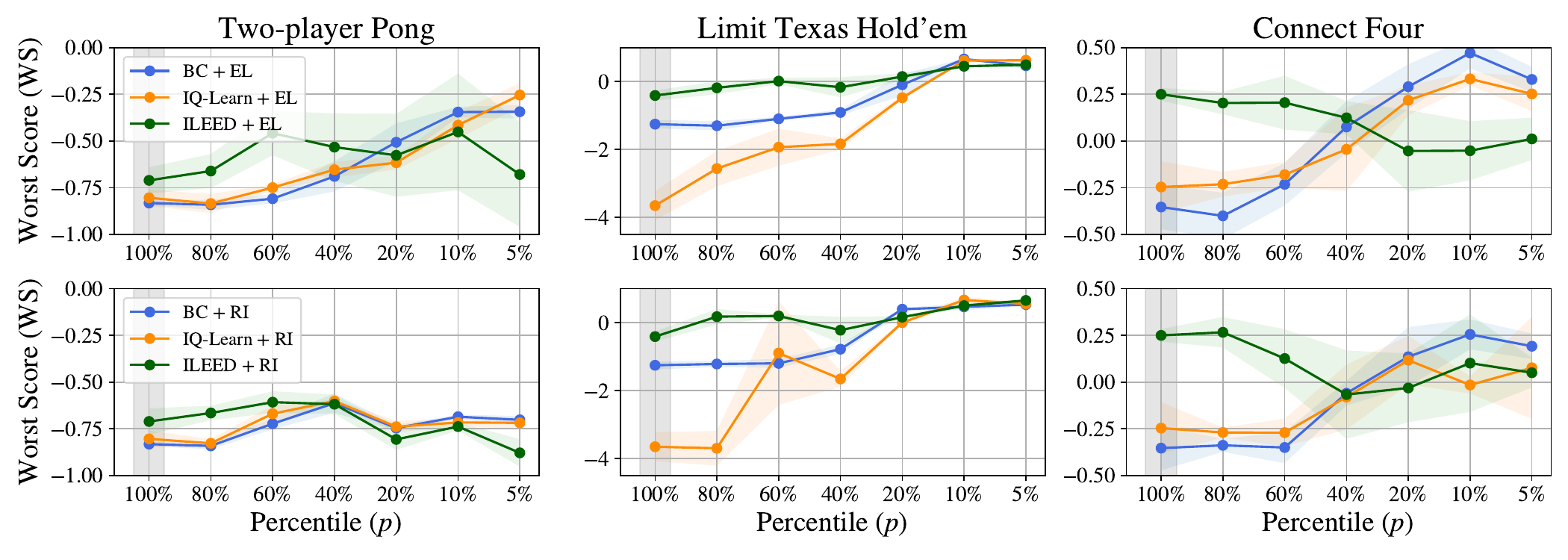}
  \vspace{-0.25cm}
  \caption{$\operatorname{WS}$ of each IL algorithm across different percentile ($p$) values for each indicator. The grey-shaded region represents the model trained on the original dataset, equivalent to the vanilla algorithm. Moving further to the right in the subfigure indicates a decrease in the data used. Higher is better.
  }
  \vspace{-0.3cm}
  \label{fig:threshold}
\end{figure*}

\textbf{Two-player Pong.}
As shown in Figure \ref{fig:pong1}, strategy representations of each demonstrator naturally cluster together in the Two-player Pong environment. Figure \ref{fig:pong4} demonstrates that trajectory rewards only partially align with player strategies due to performance variability depending on the opponent's strategy, a common characteristic of competitive games. Players $1$ and $8$, exhibiting the worst and best performances, respectively, form clusters with consistent values independent of the opponent. Figure \ref{fig:pong2} illustrates that the RI highlights players $5$, $6$, and $8$ as excelling in reconstruction tasks. Additionally, Figure \ref{fig:pong3} shows that the dominant players, $4$ and $8$, have strategies that are least susceptible to exploitation, signifying more robust performance. Additionally, it is observed that the most expansive cluster with the lowest density has the highest EL, suggesting that the least trained strategy exhibits unstable behavior and is the most vulnerable one to exploitation. These indicators establish a strong standard for data filtering in subsequent IL applications from two different perspectives, both differentiating between dominant and dominated strategies. 

\textbf{Limit Texas Hold'em \& Connect Four.} In Limit Texas Hold'em, there are seven players: two experts, three mid-level players, and two novice players. Figure \ref{fig:card1} demonstrates that the learned representations are well separated and ordered according to their expertise levels.
Figure \ref{fig:card4} illustrates that while trajectory rewards can effectively identify very poor strategies, they fail to consistently differentiate among more effective strategies, as the rewards vary across different opponents.
However, Figures \ref{fig:card2} and \ref{fig:card3} show that our proposed indicators not only perfectly distinguish the dominant strategies but also rank them accurately. In Connect Four, Figure \ref{fig:conn1} shows dominant player strategies on the right and dominated player strategies on the left.
In contrast to the trajectory rewards which are inconsistent within the same strategy, our RI and EL patterns show a strong capability to extract characteristics and assess the performance of these strategies.

\subsection{Learning from Offline Dataset}\label{sec:learn_from_off}
To evaluate the STRIL, we considered three IL algorithms. First, we employed BC, a basic IL algorithm. Next, we used IQ-Learn \cite{garg2021iq}, an advanced imitative algorithm. Finally, we implemented ILEED \cite{beliaev2022imitation}, a state-of-the-art method capable of handling a diverse range of demonstrator data.
In our evaluation, we excluded methods that rely on online interactions (e.g., GAIL \cite{ho2016generative}) or necessitate interactions with experts (e.g., DAgger \cite{ross2011reduction}) in offline learning approaches.
We applied STRIL to each algorithm to evaluate its performance enhancement.
Note that all the experiments were repeated three times.

\textbf{General results.} In Table \ref{tab:exploitability}, we compared the WS of four types of data filtering methods.
A hyperparameter search was conducted to identify the appropriate percentile, $p$, of indicators for each model and environment.
Note that all experiments were repeated three times, and the results are reported with error bars.
The original ILEED, which considers the expertise level of the data, generally performs better than other original algorithms on average. 
For the filtering method, the RI and EL enhance the performance of the original methods in most cases. In some instances, their performance is even comparable to the Best method. In the case of Two-player Pong, the EL method outperforms the RI method because EL more accurately distinguishes the dominant strategy. Additionally, in Limit Texas Hold’em, both RI and EL show similar performance, which aligns with the similar qualitative results observed in the strategy space. However, the RI and EL methods did not improve ILEED on Connect Four because the dataset aligns well with the assumption of ILEED. Consequently, filtering the data according to randomness is equivalent to reducing valid data, which results in worse performance. 

\textbf{Sensitivity analysis.} Figure \ref{fig:threshold} shows the performance for each IL algorithm across different percentile values for each indicator. For the BC and IQ-Learn algorithms, the RI and EL methods provide improved performance in all the cases. Although ILEED is designed to learn from diverse demonstrators, the RI and EL methods can be effectively used in some environments because ILEED struggles to distinguish the dominant policy in a multi-agent environment. 
For the EL method, due to the significant decrease in the size of the filtered dataset, a drop in performance from $p=0.1$ to $p=0.05$ is commonly observed. In contrast, in the range of $p\geq 0.1$, the overall performance is enhanced as $p$ decreases. EL is a reliable indicator since it has a few reward-labeled data as anchors, while RI solely takes estimated randomness as evaluation metrics. The result of the RI method across different $p$'s shows less stable behavior, as the optimal results are achieved on $p=0.4$ or $p=0.1$ in different game scenarios. However, choosing a relatively small $p$ for an unknown dataset is a preferred option since the most proficient demonstrators usually have the most stable strategies.

\section{Conclusion}\label{sec:disc}
In this work, we proposed an effective framework, STRIL, to extract the representations of the offline trajectories and enhance imitation learning methods in multi-agent games. We designed a P-VRNN network, which shows extraordinary results in learning the strategy representations of trajectories without requiring player identification. We then defined two indicators, RI and EL, for imitation learning. We can estimate RI and EL by utilizing the strategy representation and subsequently filter the offline dataset with the indicators. The imitation learning algorithms show significant performance improvements with the filtered datasets. 

In future work, we plan to utilize the P-VRNN as a customized behavior prediction model and explore the geometry of the strategy representation space. Additionally, we aim to develop IL methods that integrate the indicators beyond simply filtering the dataset.

\section{Acknowledgments}
This work was supported in part by the National Natural Science Foundation of China under Grants 72293575, as well as by the National Research Foundation of Korea (NRF) grant funded by the Korea government (MSIT) (No. RS-2024-00410082)

\bibliography{aaai25}

\begin{thebibliography}{37}
\providecommand{\natexlab}[1]{#1}

\bibitem[{Andrychowicz et~al.(2020)Andrychowicz, Baker, Chociej, Jozefowicz, McGrew, Pachocki, Petron, Plappert, Powell, Ray et~al.}]{andrychowicz2020learning}
Andrychowicz, O.~M.; Baker, B.; Chociej, M.; Jozefowicz, R.; McGrew, B.; Pachocki, J.; Petron, A.; Plappert, M.; Powell, G.; Ray, A.; et~al. 2020.
\newblock Learning dexterous in-hand manipulation.
\newblock \emph{The International Journal of Robotics Research}, 39(1): 3--20.

\bibitem[{Bansal et~al.(2018)Bansal, Pachocki, Sidor, Sutskever, and Mordatch}]{bansal2018emergent}
Bansal, T.; Pachocki, J.; Sidor, S.; Sutskever, I.; and Mordatch, I. 2018.
\newblock Emergent Complexity via Multi-Agent Competition.
\newblock In \emph{International Conference on Learning Representations}.

\bibitem[{Beliaev et~al.(2022)Beliaev, Shih, Ermon, Sadigh, and Pedarsani}]{beliaev2022imitation}
Beliaev, M.; Shih, A.; Ermon, S.; Sadigh, D.; and Pedarsani, R. 2022.
\newblock Imitation learning by estimating expertise of demonstrators.
\newblock In \emph{International Conference on Machine Learning}, 1732--1748. PMLR.

\bibitem[{Berner et~al.(2019)Berner, Brockman, Chan, Cheung, Debiak, Dennison, Farhi, Fischer, Hashme, Hesse et~al.}]{berner2019dota}
Berner, C.; Brockman, G.; Chan, B.; Cheung, V.; Debiak, P.; Dennison, C.; Farhi, D.; Fischer, Q.; Hashme, S.; Hesse, C.; et~al. 2019.
\newblock Dota 2 with large scale deep reinforcement learning.
\newblock \emph{arXiv preprint arXiv:1912.06680}.

\bibitem[{Bojarski et~al.(2016)Bojarski, Del~Testa, Dworakowski, Firner, Flepp, Goyal, Jackel, Monfort, Muller, Zhang et~al.}]{bojarski2016end}
Bojarski, M.; Del~Testa, D.; Dworakowski, D.; Firner, B.; Flepp, B.; Goyal, P.; Jackel, L.~D.; Monfort, M.; Muller, U.; Zhang, J.; et~al. 2016.
\newblock End to end learning for self-driving cars.
\newblock \emph{arXiv preprint arXiv:1604.07316}.

\bibitem[{Brown, Goo, and Niekum(2020)}]{brown2020better}
Brown, D.~S.; Goo, W.; and Niekum, S. 2020.
\newblock Better-than-demonstrator imitation learning via automatically-ranked demonstrations.
\newblock In \emph{Conference on robot learning}, 330--359. PMLR.

\bibitem[{Chen, Yuan, and Tomizuka(2019)}]{chen2019model}
Chen, J.; Yuan, B.; and Tomizuka, M. 2019.
\newblock Model-free deep reinforcement learning for urban autonomous driving.
\newblock In \emph{2019 IEEE intelligent transportation systems conference (ITSC)}, 2765--2771. IEEE.

\bibitem[{Chen, Paleja, and Gombolay(2021)}]{chen2021learning}
Chen, L.; Paleja, R.; and Gombolay, M. 2021.
\newblock Learning from suboptimal demonstration via self-supervised reward regression.
\newblock In \emph{Conference on robot learning}, 1262--1277. PMLR.

\bibitem[{Chung et~al.(2015)Chung, Kastner, Dinh, Goel, Courville, and Bengio}]{chung2015recurrent}
Chung, J.; Kastner, K.; Dinh, L.; Goel, K.; Courville, A.~C.; and Bengio, Y. 2015.
\newblock A recurrent latent variable model for sequential data.
\newblock \emph{Advances in neural information processing systems}, 28.

\bibitem[{Ding et~al.(2019)Ding, Florensa, Abbeel, and Phielipp}]{ding2019goal}
Ding, Y.; Florensa, C.; Abbeel, P.; and Phielipp, M. 2019.
\newblock Goal-conditioned imitation learning.
\newblock \emph{Advances in neural information processing systems}, 32.

\bibitem[{Fei et~al.(2020)Fei, Wang, Zhuang, Zhang, Hao, Zhang, Ji, and Liu}]{fei2020triple}
Fei, C.; Wang, B.; Zhuang, Y.; Zhang, Z.; Hao, J.; Zhang, H.; Ji, X.; and Liu, W. 2020.
\newblock Triple-GAIL: a multi-modal imitation learning framework with generative adversarial nets.
\newblock \emph{arXiv preprint arXiv:2005.10622}.

\bibitem[{Franzmeyer et~al.(2024)Franzmeyer, Elkind, Torr, Foerster, and Henriques}]{franzmeyer2024select}
Franzmeyer, T.; Elkind, E.; Torr, P.; Foerster, J.~N.; and Henriques, J.~F. 2024.
\newblock Select to Perfect: Imitating desired behavior from large multi-agent data.
\newblock In \emph{The Twelfth International Conference on Learning Representations}.

\bibitem[{Fujimoto, Meger, and Precup(2019)}]{fujimoto2019off}
Fujimoto, S.; Meger, D.; and Precup, D. 2019.
\newblock Off-policy deep reinforcement learning without exploration.
\newblock In \emph{International conference on machine learning}, 2052--2062. PMLR.

\bibitem[{Garg et~al.(2021)Garg, Chakraborty, Cundy, Song, and Ermon}]{garg2021iq}
Garg, D.; Chakraborty, S.; Cundy, C.; Song, J.; and Ermon, S. 2021.
\newblock Iq-learn: Inverse soft-q learning for imitation.
\newblock \emph{Advances in Neural Information Processing Systems}, 34: 4028--4039.

\bibitem[{Grover et~al.(2018)Grover, Al-Shedivat, Gupta, Burda, and Edwards}]{grover2018learning}
Grover, A.; Al-Shedivat, M.; Gupta, J.; Burda, Y.; and Edwards, H. 2018.
\newblock Learning policy representations in multiagent systems.
\newblock In \emph{International conference on machine learning}, 1802--1811. PMLR.

\bibitem[{Hausman et~al.(2017)Hausman, Chebotar, Schaal, Sukhatme, and Lim}]{hausman2017multi}
Hausman, K.; Chebotar, Y.; Schaal, S.; Sukhatme, G.; and Lim, J.~J. 2017.
\newblock Multi-modal imitation learning from unstructured demonstrations using generative adversarial nets.
\newblock \emph{Advances in neural information processing systems}, 30.

\bibitem[{Heinrich and Silver(2016)}]{heinrich2016deep}
Heinrich, J.; and Silver, D. 2016.
\newblock Deep reinforcement learning from self-play in imperfect-information games.
\newblock \emph{arXiv preprint arXiv:1603.01121}.

\bibitem[{Ho and Ermon(2016)}]{ho2016generative}
Ho, J.; and Ermon, S. 2016.
\newblock Generative adversarial imitation learning.
\newblock \emph{Advances in neural information processing systems}, 29.

\bibitem[{Kaelbling, Littman, and Cassandra(1998)}]{kaelbling1998planning}
Kaelbling, L.~P.; Littman, M.~L.; and Cassandra, A.~R. 1998.
\newblock Planning and acting in partially observable stochastic domains.
\newblock \emph{Artificial intelligence}, 101(1-2): 99--134.

\bibitem[{Kim et~al.(2022)Kim, Seo, Lee, Jeon, Hwang, Yang, and Kim}]{kim2022demodice}
Kim, G.-H.; Seo, S.; Lee, J.; Jeon, W.; Hwang, H.; Yang, H.; and Kim, K.-E. 2022.
\newblock Demodice: Offline imitation learning with supplementary imperfect demonstrations.
\newblock In \emph{International Conference on Learning Representations}.

\bibitem[{Kumar et~al.(2020)Kumar, Zhou, Tucker, and Levine}]{kumar2020conservative}
Kumar, A.; Zhou, A.; Tucker, G.; and Levine, S. 2020.
\newblock Conservative q-learning for offline reinforcement learning.
\newblock \emph{Advances in Neural Information Processing Systems}, 33: 1179--1191.

\bibitem[{Littman(1994)}]{littman1994markov}
Littman, M.~L. 1994.
\newblock Markov games as a framework for multi-agent reinforcement learning.
\newblock In \emph{Machine learning proceedings 1994}, 157--163. Elsevier.

\bibitem[{Lynch et~al.(2020)Lynch, Khansari, Xiao, Kumar, Tompson, Levine, and Sermanet}]{lynch2020learning}
Lynch, C.; Khansari, M.; Xiao, T.; Kumar, V.; Tompson, J.; Levine, S.; and Sermanet, P. 2020.
\newblock Learning latent plans from play.
\newblock In \emph{Conference on robot learning}, 1113--1132. PMLR.

\bibitem[{Mandlekar et~al.(2019)Mandlekar, Booher, Spero, Tung, Gupta, Zhu, Garg, Savarese, and Fei-Fei}]{mandlekar2019scaling}
Mandlekar, A.; Booher, J.; Spero, M.; Tung, A.; Gupta, A.; Zhu, Y.; Garg, A.; Savarese, S.; and Fei-Fei, L. 2019.
\newblock Scaling robot supervision to hundreds of hours with roboturk: Robotic manipulation dataset through human reasoning and dexterity.
\newblock In \emph{2019 IEEE/RSJ International Conference on Intelligent Robots and Systems (IROS)}, 1048--1055. IEEE.

\bibitem[{Mandlekar et~al.(2021)Mandlekar, Xu, Wong, Nasiriany, Wang, Kulkarni, Fei-Fei, Savarese, Zhu, and Mart{\'\i}n-Mart{\'\i}n}]{mandlekar2021matters}
Mandlekar, A.; Xu, D.; Wong, J.; Nasiriany, S.; Wang, C.; Kulkarni, R.; Fei-Fei, L.; Savarese, S.; Zhu, Y.; and Mart{\'\i}n-Mart{\'\i}n, R. 2021.
\newblock What Matters in Learning from Offline Human Demonstrations for Robot Manipulation.
\newblock In \emph{5th Annual Conference on Robot Learning}.

\bibitem[{Mnih et~al.(2013)Mnih, Kavukcuoglu, Silver, Graves, Antonoglou, Wierstra, and Riedmiller}]{mnih2013playing}
Mnih, V.; Kavukcuoglu, K.; Silver, D.; Graves, A.; Antonoglou, I.; Wierstra, D.; and Riedmiller, M. 2013.
\newblock Playing atari with deep reinforcement learning.
\newblock \emph{arXiv preprint arXiv:1312.5602}.

\bibitem[{Pomerleau(1988)}]{pomerleau1988alvinn}
Pomerleau, D.~A. 1988.
\newblock Alvinn: An autonomous land vehicle in a neural network.
\newblock \emph{Advances in neural information processing systems}, 1.

\bibitem[{Ross, Gordon, and Bagnell(2011)}]{ross2011reduction}
Ross, S.; Gordon, G.; and Bagnell, D. 2011.
\newblock A reduction of imitation learning and structured prediction to no-regret online learning.
\newblock In \emph{Proceedings of the fourteenth international conference on artificial intelligence and statistics}, 627--635. JMLR Workshop and Conference Proceedings.

\bibitem[{Sasaki and Yamashina(2020)}]{sasaki2020behavioral}
Sasaki, F.; and Yamashina, R. 2020.
\newblock Behavioral cloning from noisy demonstrations.
\newblock In \emph{International Conference on Learning Representations}.

\bibitem[{Schulman et~al.(2017)Schulman, Wolski, Dhariwal, Radford, and Klimov}]{schulman2017proximal}
Schulman, J.; Wolski, F.; Dhariwal, P.; Radford, A.; and Klimov, O. 2017.
\newblock Proximal policy optimization algorithms.
\newblock \emph{arXiv preprint arXiv:1707.06347}.

\bibitem[{Sharma et~al.(2018)Sharma, Mohan, Pinto, and Gupta}]{sharma2018multiple}
Sharma, P.; Mohan, L.; Pinto, L.; and Gupta, A. 2018.
\newblock Multiple interactions made easy (mime): Large scale demonstrations data for imitation.
\newblock In \emph{Conference on robot learning}, 906--915. PMLR.

\bibitem[{Terry et~al.(2021)Terry, Black, Grammel, Jayakumar, Hari, Sullivan, Santos, Dieffendahl, Horsch, Perez-Vicente et~al.}]{terry2021pettingzoo}
Terry, J.; Black, B.; Grammel, N.; Jayakumar, M.; Hari, A.; Sullivan, R.; Santos, L.~S.; Dieffendahl, C.; Horsch, C.; Perez-Vicente, R.; et~al. 2021.
\newblock Pettingzoo: Gym for multi-agent reinforcement learning.
\newblock \emph{Advances in Neural Information Processing Systems}, 34: 15032--15043.

\bibitem[{Vinyals et~al.(2019)Vinyals, Babuschkin, Czarnecki, Mathieu, Dudzik, Chung, Choi, Powell, Ewalds, Georgiev et~al.}]{vinyals2019grandmaster}
Vinyals, O.; Babuschkin, I.; Czarnecki, W.~M.; Mathieu, M.; Dudzik, A.; Chung, J.; Choi, D.~H.; Powell, R.; Ewalds, T.; Georgiev, P.; et~al. 2019.
\newblock Grandmaster level in StarCraft II using multi-agent reinforcement learning.
\newblock \emph{Nature}, 575(7782): 350--354.

\bibitem[{Xu et~al.(2022)Xu, Zhan, Yin, and Qin}]{xu2022discriminator}
Xu, H.; Zhan, X.; Yin, H.; and Qin, H. 2022.
\newblock Discriminator-weighted offline imitation learning from suboptimal demonstrations.
\newblock In \emph{International Conference on Machine Learning}, 24725--24742. PMLR.

\bibitem[{Yang, Levine, and Nachum(2021)}]{yang2021trail}
Yang, M.; Levine, S.; and Nachum, O. 2021.
\newblock Trail: Near-optimal imitation learning with suboptimal data.
\newblock \emph{arXiv preprint arXiv:2110.14770}.

\bibitem[{Zha et~al.(2020)Zha, Lai, Huang, Cao, Reddy, Vargas, Nguyen, Wei, Guo, and Hu}]{zha2020rlcard}
Zha, D.; Lai, K.-H.; Huang, S.; Cao, Y.; Reddy, K.; Vargas, J.; Nguyen, A.; Wei, R.; Guo, J.; and Hu, X. 2020.
\newblock RLCard: A Platform for Reinforcement Learning in Card Games.
\newblock In \emph{IJCAI}.

\bibitem[{Zhang et~al.(2021)Zhang, Cao, Sadigh, and Sui}]{zhang2021confidence}
Zhang, S.; Cao, Z.; Sadigh, D.; and Sui, Y. 2021.
\newblock Confidence-aware imitation learning from demonstrations with varying optimality.
\newblock \emph{Advances in Neural Information Processing Systems}, 34: 12340--12350.

\end{thebibliography}

\clearpage

\section{Exploited Level}\label{app:el}

\subsection{An Intuition of Exploited Level}

\begin{figure}[h]
  \centering
  \includegraphics[width=0.6\columnwidth]{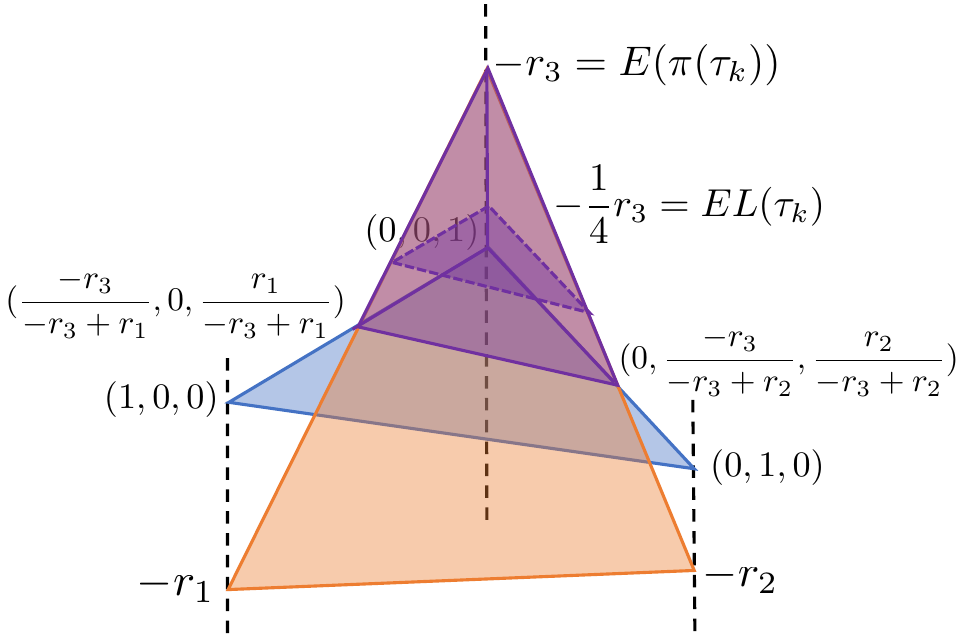}
  \caption{Illustration of EL and exploitability of a strategy in a two-player zero-sum game with three pure strategies.}
  \vspace{0.3cm}
  \label{fig:toy}
\end{figure}

In this section, we provide an intuition of \textit{Exploited Level} (EL) with a toy model. It serves as a proportional approximation of exploitability with a certain distribution on the strategy set. 
Consider a 2-player zero-sum symmetric game that has $n$ pure strategies $\xi_i$, $i=1,...,n$. 
All strategies are convex combinations of pure strategies, i.e., $\pi =\sum_{i=1}^{n}\alpha_i \xi_i$, where $\sum_{i=1}^{n}\alpha_i=1$, $0\leq \alpha_i \leq 1, \forall i$.
For simplicity, we assume that each trajectory $\tau$ can be directly mapped to a strategy $\pi(\tau)$. 
In our setting, where all the players are competent, each player can be exploited by at most one pure strategy. 
As for the overall strategy distribution over $\Pi$ (the strategy space), we assume the $(\alpha_1, \alpha_2, ..., \alpha_n)$ has uniform distribution over $(n-1)$-dimensional standard simplex.

The definition of EL is as follows:
\begin{equation*}
EL(\tau)=\mathbb{E}_{\pi}\left[-r(\pi,\pi(\tau))\mid r(\pi,\pi(\tau))\leq 0\right].
\end{equation*}
For a trajectory $\tau_k$, let $r(\xi_i,\pi(\tau_k))=r_i$. By our assumption, only one $j\in\{1,2,...,n\}$ satisfies that $r_j<0$, while $r_i\geq 0, \forall i\neq j$. We can directly see that $E(\pi(\tau_k))=-r_j$. 

Since $EL(\tau_k)$ is a conditional expectation defined on $\Pi$, we can view it as a conditional expected value over an $(n-1)$-dimensional simplex. 
When $\pi =\sum_{i=1}^{n}\alpha_i \xi_i$, $r(\pi,\pi(\tau_k))=\sum_{i=1}^{n}\alpha_i r_i$, the condition becomes $\sum_{i=1}^{n}r_i \alpha_i\leq 0$. 
Thus, the expectation is still defined over an $(n-1)$-dimensional simplex, but a smaller one, with vertices $\{(0,...,\alpha_i=\frac{-r_j}{-r_j+r_i},...,\alpha_j=\frac{r_i}{-r_j+r_i},...,0)\mid \forall i\neq j\}\cup \{(0,...,\alpha_j=1,...,0)\}$. 
Then, we can consider adding another dimension on the simplex, so that the new dimension has value $-r(\pi,\pi(\tau))$. Due to linearity, the new object becomes an $n$-dimensional pyramid, and the desired expectation is the height of the pyramid's centroid w.r.t. the surface of the original $(n-1)$-dimensional simplex. 
From calculus, the height of the centroid of $n$-dimensional pyramid is always $\frac{1}{n+1}$ of the height of the pyramid w.r.t. its base. Since the height is $r_j$, the expectation is $\frac{1}{n+1}r_j$. 
So 
\begin{equation*}
EL(\tau_k)=\frac{1}{n+1}E(\pi(\tau_k))
\end{equation*}
always holds in this case, which shows that EL is an appropriate indicator. A strategy of a game with three different pure strategies is shown as an example in Figure \ref{fig:toy}, with EL and exploitability visualized.

Concretely, consider an RPS game and let $\xi_1,\xi_2$ and $\xi_3$ be the pure strategies of choosing rock, paper, and scissors, respectively. Let the strategy of $\tau$ be $\pi(\tau)=(0,2/3,1/3)$, i.e. "choosing paper with $2/3$ probability and choosing scissors with $1/3$ probability". Then we can easily derive that $-r_1=-r_2=-1/3, -r_3=2/3$. So we have $E(\pi(\tau))=2/3$, while $EL(\tau)=1/6$.

\subsection{Why Exploited Level?}\label{subsec:ele}

In two-player symmetric zero-sum games, it is common to use exploitability as a measure for evaluating the effectiveness of a strategy. 
However, it is extremely difficult to obtain exploitability with a single trajectory since we cannot: 1) infer or modify the strategy of the opponent; or 2) make any interaction with the environment.
For a strategy $\pi_i$, if we have many trajectories that have a strategy similar to it and the opponents use a large variety of strategies (so that there is one strategy near the best response), then $\forall \epsilon>0$, there exists a $\delta>0$ which satisfies the following approximation:
\begin{equation*}
\left|E(\pi_i)-\max_{d(\pi_i',\pi_i)<\delta}\left[-r(\hat{\pi}_{-i},\pi_i')\right]\right|<\epsilon,
\end{equation*}
where $d$ is a distance over the strategy space.

However, if we have many trajectories so that for each trajectory, the opponent strategies can cover most kinds of strategies, and the trajectories with similar representation vectors have similar strategy distributions, can we still use the minimum reward of trajectories with representation near itself to serve as an approximation of negative exploitability? 
First, we define measure $\text{d}\pi$ on strategy space $\Pi$ according to the probability of $\pi$ chosen in the whole dataset:
\begin{equation*}
\int_{\pi\in S}\text{d}\pi=\mathbb{P}\left[\tau \sim \pi, \pi \in S,\forall\tau\in\Gamma\right],
\end{equation*}
where $S$ is an arbitrary subset of $\Pi$.
Denote the trajectory as $\tau$, the representation function learned above as $f(\tau)$, and the reward of $\tau$ as $r(\tau)$. 
We remark that a trajectory $\tau$ should be mapped to a probability distribution of strategies such that $\int_{\pi\in\Pi}\tau(\pi)\text{d}\pi=1$, where $\tau(\pi)$ is the probability of using strategy $\pi$ when having trajectory $\tau$, instead of a single strategy. But we can view the mixture of $\pi$ with probability $\tau(\pi)$ as a single mixed strategy $\int_{\pi\in\Pi}\pi\tau(\pi)\text{d}\pi$, so we can still use notation $\pi(\tau)$ to represent the strategy of $\tau$. 
Using the above method, we can approximate $E(\pi(\tau))$, i.e.,
\begin{equation*}\left|E(\pi(\tau))-\max_{d(f(\tau'),f(\tau))<\delta}\left[-r(\tau')\right]\right|<\epsilon.\end{equation*}
But the $E(\pi(\tau))$ we are approximating is \textbf{not} what we desire. 
In order to measure the exploitability of $\tau$, we should calculate $E(\tau):=\int_{\pi\in\Pi}\tau(\pi)E(\pi)\text{d}\pi$ instead of $E\left(\int_{\pi\in\Pi}\pi\tau(\pi)\text{d}\pi\right)$. 
We have the following result:
\begin{proposition}
    If $\tau(\pi)$ is a distribution over $\Pi$, and $E$ is defined as exploitability, then we have
    \begin{equation*}\int_{\pi\in\Pi}\tau(\pi)E(\pi)\text{d}\pi\geq E\left(\int_{\pi\in\Pi}\pi\tau(\pi)\text{d}\pi\right).
    \end{equation*}
    \label{prop1}
\end{proposition}

Given the proposition above, there will be an underestimation if we use this method. Also, using maximum alone abandons almost all the information of nearby trajectories, which makes the approximation unstable.
To resolve these problems, we use mean instead of maximum. 
Here, we restate the definition of the exploited level (EL) as
\begin{equation*}
EL(\tau)=\mathbb{E}_{\pi}\left[-r(\pi,\pi(\tau))\mid r(\pi,\pi(\tau))\leq 0\right].
\end{equation*}
Except for the conditions mentioned above, the algorithm is mainly based on the following assumption: 
\begin{equation*}
E(\tau)\propto EL(\tau)= \frac{\int_{\pi\in\Pi}(-r(\pi,\pi(\tau))^{+}\text{d}\pi}{\int_{\pi\in\Pi}\mathds{1}_{r(\pi,\pi(\tau))\leq 0}\text{d}\pi},
\end{equation*}
where $r(\pi,\pi(\tau))$ returns the reward of a player with strategy $\pi(\tau)$ by default, $(x)^+=\max{\{x,0\}}$ and $\mathds{1}_{c}=1$ if and only if condition c is satisfied, otherwise $\mathds{1}_{c}=0$. 
The above function means that given a trajectory $\tau$, the mean negative reward of the trajectories with a representation near $\tau$ and reward less than $0$ is proportional to exploitability. 
The right-hand side value is a reasonable measure of a trajectory, which is shown in the toy model. To estimate EL with latent representation space, we provide an alternative definition of $EL_\delta$: 
\begin{equation*}
EL_\delta(\tau)=\frac{\sum_{d(f(\tau),f(\tau'))<\delta}(-r(\hat{\pi}, \pi(\tau')))^+}{\sum_{d(f(\tau),f(\tau'))<\delta}\mathds{1}_{r(\hat{\pi}, \pi(\tau'))\leq 0}}.
\end{equation*}
It is obvious that $\lim_{\delta\to 0^{+}}EL_\delta(\tau)=EL(\tau)$.
The property of EL satisfies our requirement that the trajectories that perform similarly to Nash Equilibrium can be detected with an EL near $0$ since we have the following proposition.
\begin{proposition}\label{prop2}
    Given a trajectory $\tau$ and its corresponding distribution $\tau(\pi)$ over $\Pi$, $\pi(\tau)$ is $\epsilon_1$-Nash equilibrium, and we assume that any pure strategy can exploit another strategy by at most $M$. 
    By the smoothness of $f$, we also assume that if $d(f(\tau_1),f(\tau_2))<\delta$, then $\int_{\pi\in\Pi}|\tau_1(\pi)-\tau_2(\pi)|\text{d}\pi<\alpha\delta$, where $\alpha$ is a constant. We have the following result:
    \begin{equation*}
    EL_\delta(\tau)<\epsilon_1+\alpha\delta M.
    \end{equation*} 
\end{proposition}

\section{The Proof of Proposition \ref{prop1}}\label{app:proof1}

\begin{proof}
    For simplicity, we only prove in a 2-player setting. By definition of exploitability, $E(\pi)=-r(BR(\pi),\pi)$. So we have
    \begin{align*}
        E(\pi(\tau)) &= -r\left(BR(\pi(\tau)),\pi(\tau)\right) \\
        &= -r\left(\text{argmax}_{\pi_{-i}}r(\pi_{-i},\pi(\tau)), \pi(\tau)\right) \\
        &= -\int_{\pi\in\Pi}\tau(\pi)r\left(\text{argmax}_{\pi_{-i}}r(\pi_{-i},\pi(\tau)),\pi\right)\text{d}\pi \\
        &\leq -\int_{\pi\in\Pi}\tau(\pi)r\left(\text{argmax}_{\pi_{-i}}r(\pi_{-i},\pi),\pi\right)\text{d}\pi \\
        &= -\int_{\pi\in\Pi}\tau(\pi)r\left(BR(\pi),\pi\right)\text{d}\pi \\
        &= \int_{\pi\in\Pi}\tau(\pi)E(\pi)\text{d}\pi
    \end{align*}
    The inequality is established by the property of $\text{argmax}$ function.
\end{proof}

\section{The Proof of Proposition \ref{prop2}}\label{app:proof2}

\begin{proof}
    Since $\pi(\tau)$ is $\epsilon_1$-Nash equilibrium, the exploitability $E(\pi(\tau))\leq \epsilon_1$. Thus for an arbitrary $\hat{\pi}$, we have $r(\hat{\pi},\pi(\tau))\geq -\epsilon_1$. Hence, for all $\tau'$ satisfying $d(f(\tau),f(\tau'))<\delta$, we have
    \begin{align*}
        r(\hat{\pi},\pi(\tau')) &= \int_{\pi\in\Pi}\tau'(\pi)r(\hat{\pi},\pi)\text{d}\pi \\
        &= \int_{\pi\in\Pi}\tau(\pi)r(\hat{\pi},\pi)\text{d}\pi + \int_{\pi\in\Pi}(\tau'(\pi)-\tau(\pi))r(\hat{\pi},\pi)\text{d}\pi \\
        &\geq r(\hat{\pi},\pi(\tau))-\int_{\pi\in\Pi}\left|\tau'(\pi)-\tau(\pi)\right|\left|r(\hat{\pi},\pi)\right|\text{d}\pi \\
        &\geq -\epsilon_1-M\int_{\pi\in\Pi}\left|\tau'(\pi)-\tau(\pi)\right|\text{d}\pi \\
        &> -\epsilon_1-\alpha\delta M
    \end{align*}
    Thus, we have
    \begin{align*}
        EL_\delta(\tau) &= \frac{\sum_{d(f(\tau),f(\tau'))<\delta}(-r(\hat{\pi}, \pi(\tau')))^+}{\sum_{d(f(\tau),f(\tau'))<\delta}\mathds{1}_{r(\hat{\pi}, \pi(\tau'))\leq 0}} \\
        &\leq \max_{d(f(\tau),f(\tau'))<\delta}-r(\hat{\pi}, \pi(\tau')) \\
        &< \epsilon_1+\alpha\delta M
    \end{align*}
\end{proof}

\section{The Games and Implementation Details}\label{app:imple}
\subsection{Overview of the Zero-Sum Games}\label{app:games}
We choose the following well-known games in our experiments: 
\begin{itemize}
    \item \textbf{Rock-Paper-Scissors (RPS)}: Players have three potential actions to take: rock, paper, and scissors. The observation of each player is the action of the opponent in the last round. In each trajectory, RPS games are played for $T=500$ times consecutively. The player who wins gets $+1$ point, and the player who loses gets $-1$ point. When there is a draw, the point is not changed.
    \item \textbf{Two-player Pong}: Each player controls a paddle on one side of the screen. The goal is to keep the ball in play by moving the paddles up or down to hit it. If a player misses hitting the ball with their paddle, it loses the game. The observation of players includes ball and paddle positions across two consecutive time steps and potential actions include moving up or down.
    \item \textbf{Limit Texas Hold'em}: Players start with two private hole cards, and five community cards are revealed in each stage (the flop, turn, and river). Each player has to create the best five-card hand using a combination of their hole and the community cards. During the four rounds, players can select call, check, raise, or fold. The players aim to win the game by accumulating chips through strategic betting and building strong poker hands. The observation of players is a 72-element vector, with the first 52 elements representing cards (hole cards and community cards) and the last 20 elements tracking the betting history in four rounds.
    \item \textbf{Connect Four}: Connect Four is a two-player game where the goal is to connect four of your tokens in a row, vertically, horizontally, or diagonally. The game is played on a grid with seven columns and six rows. Players drop tokens into the columns, each falling to the lowest available spot. A column can not be used if it is full. The game ends when a player connects four tokens in a row or when all columns are filled, resulting in a draw. The game state is represented by an 84-element vector, showing whether each cell has Player 1's or Player 2's token. Because Connect Four is a turn-based game, it is a perfect information game.
\end{itemize}

\subsection{Details of Generating Dataset}\label{app:details_dataset}

\begin{table}
\centering
\resizebox{1.\linewidth}{!}{
\begin{tabular}{cccc}
\toprule
& Two-player Pong& Limit Texas Hold'em& Connect Four\\
\toprule
$|\,o\,|$ & 8& 72& 84\\
$|\,a\,|$ & 2& 5& 7\\
$N_\text{dem}$ & 8& 7& 7\\
$|\,\Gamma\,|$ & 64K & 49K & 49K\\
$T$ & 500 & 100 & 200\\
\bottomrule
\end{tabular}
}
\caption{Parameters of Dataset}
\label{table:parameter_data}
\end{table}

As described in experiment settings, we trained expert policies using self-play with opponent sampling for Two-player Pong and Connect Four and neural fictitious self-play with DQN for Limit Texas Hold'em. We selected $N_{\text{dem}}$ behavior models from various intermediate checkpoints. These selected models played against each other in all possible combinations. For each pair, we generated 10K trajectories with a length of $T$, depending on the game duration. The details are presented in Table \ref{table:parameter_data}. Consequently, our offline dataset consists of $10 \times N_{\text{dem}}^{2}$ trajectories.

\subsection{Details of P-VRNN Implementation}\label{app:details_pvrnn}
In the actual implementation of P-VRNN, the action $a_t$ and observation $o_t$ pass through neural networks $\psi_\text{a}$ and $\psi_\text{o}$ first to reduce dimension and extract features.
The functions $\phi_{\text{p}}$, $\phi_{\text{e}}$, and $\phi_{\text{d}}$ are implemented with multi-layer perceptron (MLP) with latent space dimension $z_\text{dim}=8$, hidden layer dimension $h_\text{dim}=32$, recurrence layer dimension $r_\text{dim}=32$ and representation dimension $l_\text{dim}=8$ for the Two-player Pong. We set $z_\text{dim}=2$ and $l_\text{dim}=2$ for the other environments. Gated Recurrent Unit (GRU) is used as the recurrence function $\phi_{\text{r}}$. We trained the models for $500$ epochs with a learning rate of $0.001$ and a batch size of $128$ trajectories using the Adam optimizer.

\subsection{Details of Imitation Learning Experiments}\label{app:details_imitation}
In our offline learning experiments, we utilize an MLP architecture for the actor network, with two hidden layers of $256$ units each. During offline learning, we trained the models for $500$ epochs with a learning rate of $0.0001$. We set the minibatch number to $50$ for each epoch, employing the Adam optimizer to ensure a consistent number of updates for all methods. We used an official codebase for IQ-Learn\footnote{https://github.com/Div99/IQ-Learn} and ILEED\footnote{https://github.com/Stanford-ILIAD/ILEED} to ensure consistency and reproducibility.
All experiments were conducted using an RTX 2080 Ti GPU and an AMD Ryzen Threadripper 3970X CPU.

\section{Generated Offline Dataset Analysis}\label{app:data}

\subsection{Cross-Evaluation Results}

\begin{figure}[t!]
  \centering
  \subfloat[Two-player Pong]
  {
  \label{fig:cross_eval_pong}\includegraphics[width=1.\columnwidth]{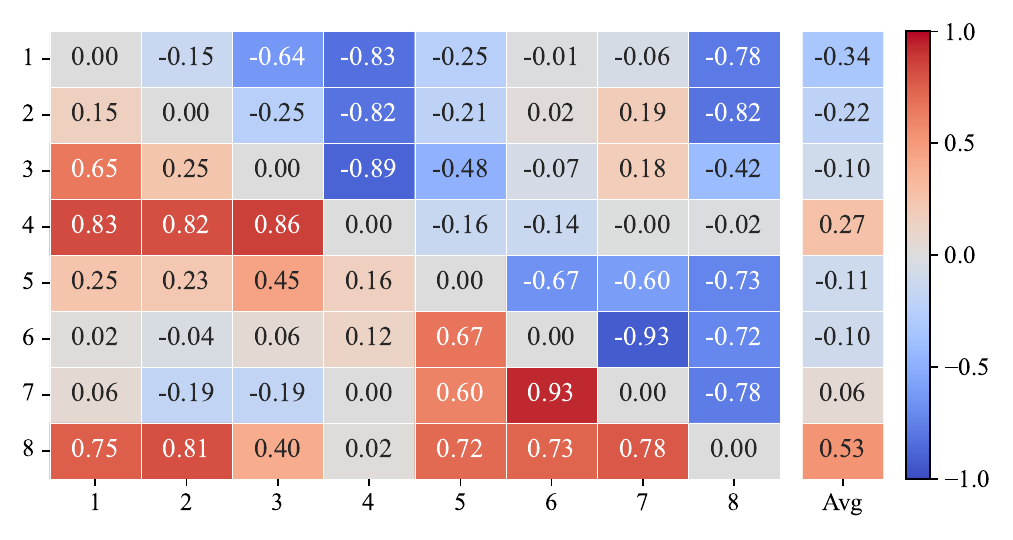}
  }\\
  \subfloat[Limit Texas Hold’em]
  {
    \label{fig:cross_eval_card}\includegraphics[width=1.\columnwidth]{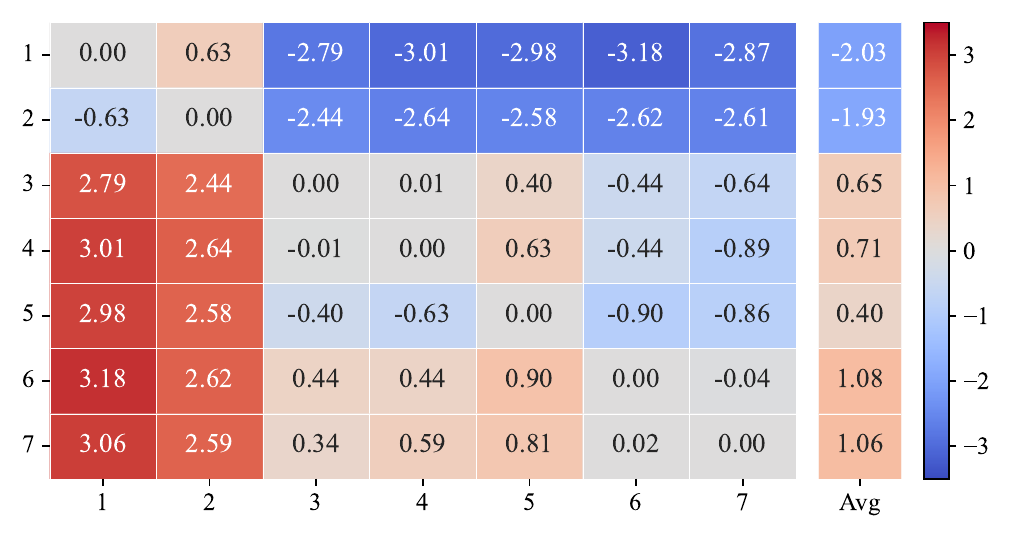}
  }\\
  \subfloat[Connect Four]
  {
  \label{fig:cross_eval_connectfour}\includegraphics[width=1.\columnwidth]{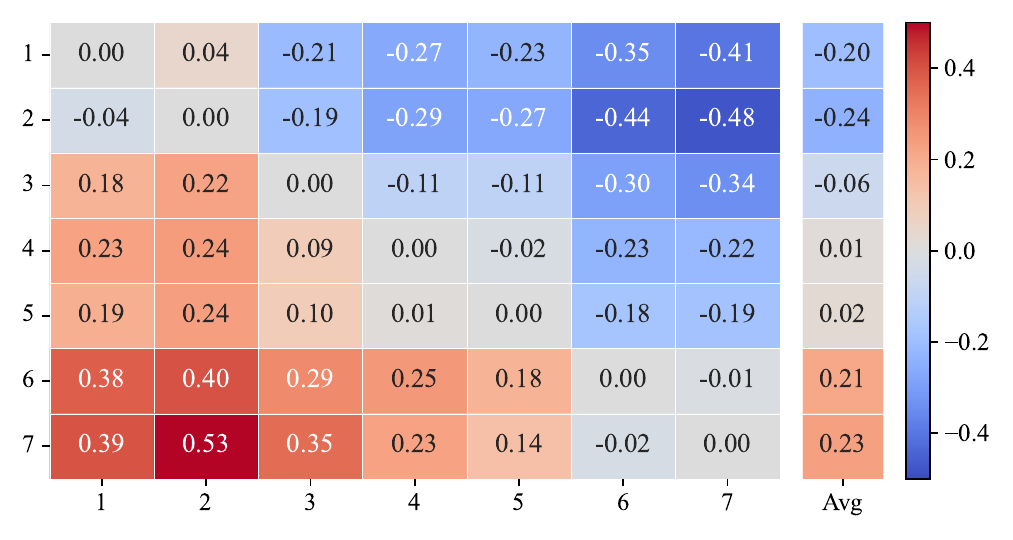}
  }
  \caption{Cross-evaluation of demonstrators in multiple games. Higher is better.}
  \label{fig:cross_Eval}
\end{figure}

We provide the cross-evaluation of demonstrators in all three games tested in Figure \ref{fig:cross_Eval}. The result reflects the diversity of overall performance and complex relationships among the demonstrators. 

\subsection{Entropy of Strategies}

We provide the average entropy of strategies on the sampled trajectories $\tau_1,...,\tau_N$, defined as
\begin{equation*}
    H(\pi)=\frac{1}{N}\sum_{i=1}^N \frac{1}{|\tau_i|}\sum_{s\in \tau_i}\sum_{a\in A_k}-\pi(a|s)\log \pi(a|s),
\end{equation*}
where $A_k$ is the action space of corresponding player $k$. We provide the results for the player strategies of Two-Player Pong and Connect Four to illustrate the difference in the strategy sets used for generating the offline dataset of the two games. We sample $N=5$ trajectories for each entropy calculation.

\begin{figure}[t]
  \centering
  \subfloat[Two-Player Pong]
  {
  \label{fig:entropy_pong}\includegraphics[width=0.4\columnwidth]{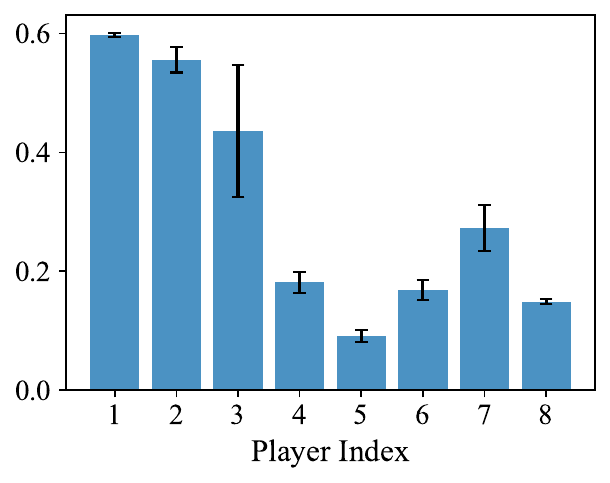}
  }
  \subfloat[Connect Four]
  {
    \label{fig:entropy_connect}\includegraphics[width=0.4\columnwidth]{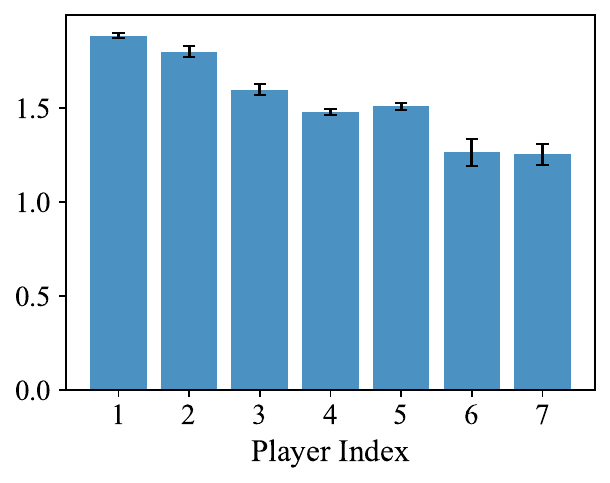}
  }
  \caption{Entropy of demonstrator strategies generating the offline datasets.}
\end{figure}

In Two-Player Pong, as shown in Figure \ref{fig:entropy_pong}, the strategy of Player 5 has the lowest entropy. However, this player is heavily exploited by Players 6, 7, and 8, as visualized in Figure \ref{fig:cross_eval_pong}. Listing out the inverse ranking of entropy and the ranking of cross-evaluation results, they are $(5,8,6,4,7,3,2,1)$ and $(8,4,7,3,6,5,2,1)$, respectively. The rankings do not match well with each other, thus providing a space for improving the performance of ILEED. 
Conversely, as for Connect Four, the entropy of the demonstrators reflects the cross-evaluation result accurately, as shown in Figure \ref{fig:entropy_connect} and Figure \ref{fig:cross_eval_connectfour}. Given the observation above, the generated offline dataset matches the assumption of ILEED well, thus it is hard for RI and EL to enhance its performance on this dataset.

\balance

\section{Limitations}\label{app:limit}
The limitations of our work emerge when the offline datasets have undesirable properties for specific indicators.
As for the Randomness Indicators, the estimation fails when the demonstrators of offline trajectories only adopt deterministic but poor strategies. 
As for the Exploited Levels, the estimation fails when the sampled trajectories with rewards are biased, covering a small area of representation space or providing biased choices of demonstrators. 
The limitations can be mitigated with a larger dataset or provided with the ability of online interactions.

\section{Broader Impacts}\label{app:social_impacts}
Our paper introduces a novel approach to learning the strategy representations and indicators for trajectories in multi-agent games. The improved efficiency in identifying dominant strategies may inadvertently amplify strategic advantages in competitive domains, posing risks to fairness. Ethical considerations are necessary to responsibly deploy the method and mitigate potential negative results in real-world applications.

\end{document}